\documentclass[aip,nofootinbib,12pt]{revtex4-1}
\usepackage{amssymb,amsmath,amsthm}
\usepackage{textcomp}
\usepackage{multido}
\usepackage{braket}
\usepackage{multido}
\usepackage{color}
\usepackage{soul}
\usepackage{latexsym}
\usepackage[english]{babel} 
\usepackage[latin1]{inputenc}  
\usepackage[T1]{fontenc}   
\usepackage[all]{xy}
\usepackage{amsfonts}
\usepackage[all]{xy}
\usepackage[colorlinks=true,linkcolor=blue]{hyperref}%

\def\CC{\mathbb{C}}

\def\C{\mathbb{C}}

\def\PP{\mathbb{P}}

\usepackage{tikz}
\usetikzlibrary{automata}
\usetikzlibrary{shadows}
\usetikzlibrary{arrows}
\usetikzlibrary{shapes}
\usetikzlibrary{fit}
\usetikzlibrary{matrix}
\newcommand{\incl}{\ar@{^{}-}}
\newcommand{\inclu}{\ar@{^{}.}}

\def\S{{\mathfrak  S}}

\newtheorem{prop}{Proposition}[section]

\newtheorem{def }{Definition  }[section]
\newtheorem{theorem}{Theorem}
\newtheorem{ex}{Example     }[section]
\newtheorem{rem}{Remark    }[section]

\newcommand{\ds}{\displaystyle}
\def\SLOCC{\mathrm{SLOCC}}

\begin{document}

 \title{Entanglement of four-qubit systems: a geometric atlas with polynomial compass  II (the tame world)}
\author{Fr\'ed\'eric Holweck\footnote{frederic.holweck@utbm.fr, 
IRTES-UTBM, Universit\'e de Bourgogne Franche-Comt\'e, 90010 Belfort Cedex, FR}, 
Jean-Gabriel Luque\footnote{jean-gabriel.luque@univ-rouen.fr, Universit\'e de Rouen, Laboratoire d'Informatique, du Traitement de l'Information et des Syst\`emes (LITIS), Avenue de l'Universit\'e - BP 8
6801 Saint-\'etienne-du-Rouvray Cedex, FR } and Jean-Yves Thibon\footnote{jyt@univ-mlv.fr, \ Laboratoire d'Informatique Gaspard Monge
Universit\'e Paris-Est Marne-la-Vall\'ee,
77454 Marne-la-Vall\'ee Cedex 2, FR}}

 \begin{abstract}
We propose a new approach to the geometry of the four-qubit entanglement classes depending on parameters.
More precisely, we use invariant theory and algebraic geometry to describe various stratifications of the Hilbert space by SLOCC invariant algebraic varieties.
The normal forms of the four-qubit classification of Verstraete {\em et al.} are interpreted as dense subsets of components of the dual variety of the set of separable states and 
an algorithm based on the invariants/covariants of the four-qubit quantum states is proposed to identify a state with a SLOCC equivalent normal form (up to qubits permutation).
 \end{abstract}

\maketitle

\section{Introduction}
Entanglement is nowdays considered as a central ressource in quantum information processing. 
The large amount of work produced since the beginning of the $\text{XXI}^{\text{st}}$ century to understand the nature of entanglement 
 demonstrates the interest of the community in this subject (see the review papers\cite{HHHH,ES} and the references therein). The question of the classification of entanglement of pure multipartite 
  quantum states 
 under the group SLOCC (Stochastic Local Operations with Classical Communication) is one of the most
prominent challenging problems on the road to the understanding of  entanglement.
The complexity of this question grows up exponentially with the number of 
parts composing the quantum system. If the classification  problem is completely solved for a few cases, it is known to be out of reach for most general situation (see Ref\cite{LT5} for a discussion on the 
 computational complexity of the algebraic invariants 
 of five-qubit systems).
 
 The case of the four-qubit Hilbert space is of special importance. First it has generated a large number of papers by itself\cite{VDMV,CD, CDGZ,BDD,Levay1, Levay3, LT, My,GW}. Then, as long as qubits are concerned, 
 it is the only case where the number 
 of SLOCC orbits is infinite but for which there still exists a list of 9 normal forms found by Verstraete {\em et al.}\cite{VDMV,CD} (6 of them depend on parameters) which, up to permutation of the qubits, parametrize all SLOCC orbits 
 (i.e. up to permutation of the qubits, a given state is in one of the orbit of the 9 families). The four-qubit case also showed up in different contexts like in the study of the black-holes-qubits correspondence\cite{BDL}, in the study of graph-states\cite{GJMW} and in many quantum communication protocols like error correcting codes\cite{GBP}.
 Thus we consider the four-qubit case as rich enough to motivate 
 more investigation and develop new tools which we hope to be useful to study more difficult cases.

 This paper is part of a sequence of articles\cite{HLT, HLT2} where we have  investigated the structure of entanglement for small multipartite systems by combining two different approches: on one side we consider classical invariant theory and 
 we try to understand what the invariants, covariants of the multipartite systems tell us about the SLOCC-orbits and, on the other side, we look at the geometrical structure of the space by building SLOCC algebraic varieties.
 Combining the two approaches, we obtain a stratification of the Hilbert space by SLOCC algebraic varieties (closure of classes or of union of classes) 
 and criteria based on invariants/covariants, to distinguish them. In our first paper\cite{HLT} we  provided a geometric description of all entanglement classes with invariants/covariants criteria
 for Hilbert spaces with a finite number of orbits. In Ref\cite{HLT2} we started to consider the four-qubit case by first looking at specific subvarieties of this Hilbert space.
 The first subvariety we considered was the nullcone, i.e. the variety of states which annihilate all SLOCC invariant polynomials. It turns out that the number of orbits contained in 
 this variety is finite and thus the techniques developped in Ref\cite{HLT} could apply, leading to a  classification, of the nilpotent four-qubit states, using covariants.
 In the same paper we also considered the case of the algebraic variety $\sigma_3(X)$, the third secant variety (see below for the definitions), defined by the simultaneous vanishing of $L$ and $M$, the 
 two degree four 
 generators of the ring of SLOCC invariant polynomials (see below for the definition). This latter case already contains an infinite number of orbits.
 
 In this paper we investigate the geometry of the four-qubit Hilbert space by considering specific SLOCC invariant hypersurfaces: the hypersurfaces defined by the vanishing of the invariants of degree four
 $L, M$ and $N$ and the so-called hyperdeterminant\cite{GKZ} $\Delta$ of format $2\times 2\times 2\times 2$. The idea of studing the geometry of the hypersurface defined by $\Delta$ 
  to classify entanglement classes appeared
 ten years ago in the work of Miyake\cite{My}. The paper of Miyake is based on the study of the singular locus of the hypersurface described ten years earlier by Weymann and Zelevinsky\cite{WZ}.
 In this article we go further in the description of the stratification of the ambient space by singular locus of the hyperdeterminant. We consider singularities leading to deeper stratas 
 and 
 we are able to establish a connection between the statas of the ambient space defined by the singularities of $\Delta$ and the $9$ families given by Verstraete {\em et al}.'s classification. 
 Like in our previous papers\cite{HLT,HLT2} the geometric approach is combined with a classical invariant theory point of view allowing us to identify the strata to 
 which  a given state belongs. It leads to an algorithm based on invariants 
 and covariants which provides, up to a qubit permutation, the family and parameters of a representative SLOCC equivalent to a given state.
 
 The paper is organized as follow. In Section \ref{adjoint} we illustrate how the adjoint representation of $SO(8)$ on its Lie algebra $\mathfrak{so}(8)$ is connected to 
 the SLOCC ($=SL_2(\CC)^{\times 4}$) orbits of $\mathcal{H}=\CC^2\times\CC^2\times\CC^2\times \CC^2$ by showing that the generators of the ring of  invariants of four-qubit can be obtained by restriction of the generators of 
 the $SO(8)$ polynomial invariants on $\mathfrak{so}(8)$. Similarly we show that the discriminant of the adjoint representation of $SO(8)$, also known as the defining equation of the dual of the adjoint variety, 
 leads to the $2\times 2\times 2\times 2$-hyperdeterminant. In the proof, the hyperdeterminant is  obtained as the discriminant of a quartic. This quartic depends on the embedding of 
 $\mathcal{H}$ in $\mathfrak{so}(8)$. In Section \ref{invariants} we investigate the three quartics corresponding to the three natural embeddings of $\mathcal{H}$ in $\mathfrak{so}(8)$.
 It leads to a first stratification of the ambient space based on the roots of the quartics. In particular, we show that this stratification provides  a diagram of normal forms which is connected to the geometry of a special semiregular polytope: the demitesseract.
 In Section \ref{geo} the geometry of SLOCC hypersurfaces corresponding to specific dual varieties is investigated. 
 This section ends with a geometric stratification of the ambient space where 
 the families of the four-qubit classification are in correspondence  with geometric stratas. Moreover some of those stratas are given by a simple geometric interpretation in terms of duals of orbits
 of well known quantum states.
 Finally, combining Section \ref{invariants} and \ref{geo} we give an algorithm based on invariants and covariants of four-qubit to determine the Verstraete type of a given state.
 Section \ref{conclu} is dedicated to  concluding remarks.
 
 \par{\bf Notations}\\
All along the paper a four-qubit state $\ket{\varphi}$ will be denoted by $\ket{\varphi}=\sum_{i_1,\dots,i_4\in\{0,1\}^4} a_{i_1\dots i_4}\ket{i_1\dotsi_4}$ 
where $\ket{i_1\dotsi_4}$ stand for the vectors 
of the computational basis. When not specified the Hilbert space $\mathcal{H}$ will be the space of pure four-qubit quantum states, i.e. $\mathcal{H}=\CC^2\otimes\CC^2\otimes\CC^2\otimes \CC^2$ and 
the corresponding SLOCC group will be SLOCC=$SL_2(\CC)^{\times 4}$.
When we work over the projective space $\PP(\mathcal{H})$, an algebraic variety $Z\subset \PP(\mathcal{H})$ is defined as the zero
 locus of a collection of homogeneous polynomials. For any subset $Y\subset \PP(V)$, 
the notation $\overline{Y}$ will refer to the Zariski closure\cite{Ha}.

\section{Adjoint variety of $SO(8)$}\label{adjoint}
The classification of four-qubit pure states under SLOCC is strongly connected to the Lie algebra $\mathfrak{so}_8$. In the original paper of Verstraete {\em et al.}\cite{VDMV}, 
the action of  the $SO(4)\times SO(4)$ subgroup of $SO(8)$, 
on the space of $8\times 8$ matrices is used to provide a first classification of SLOCC orbits of $\mathcal{H}=\CC^2\otimes\CC^2\otimes \CC^2\otimes \CC^2$. This idea is made more precise, and the classification corrected, in the 
work of Chterental and Djokovi\'c \cite{CD} where the Lie algebra $\mathfrak{so}_8$ is decomposed under the involution defined by the diagonal matrix $g=\begin{pmatrix}
                                                                                                                                                          I_4 & 0\\
                                                                                                                                                          0 & -I_4
                                                                                                                                                         \end{pmatrix}$:
                                                                                                                                                         
\begin{equation}
 \mathfrak{so}_8=\mathfrak{k}\oplus\mathfrak{p}.
\end{equation}

The $SO(4)\times SO(4)$ invariant subspace $\mathfrak{p}$ is  \begin{equation}\mathfrak{p}=\{\begin{pmatrix}
                      0 & R\\
                      -R^T & 0
                     \end{pmatrix}, R\in \mathcal{M}_4(\CC)\}\end{equation} 
                     The action of $SO(4)\times SO(4)$ on $\mathfrak{p}$ corresponds to the action of SLOCC on $\mathcal{H}=\CC^2\otimes\CC^2\otimes\CC^2\otimes\CC^2$. The correspondence is given as follows.
                     Let $\ket{\varphi}=\sum_{i,j,k,l\in \{0,1\}} a_{ijkl}\ket{ijkl}$ be a four-qubit state. One associates with $\ket{\varphi}$ the $4\times 4$ matrix
                     \begin{equation}\label{embedd}
                      M_\varphi=\begin{pmatrix}
                              a_{0000}&a_{0010}&a_{0001}&a_{0011}\\
a_{1000}&a_{1010}&a_{1001}&a_{1011}\\
a_{0100}&a_{0110}&a_{0101}&a_{0111}\\
a_{1100}&a_{1110}&a_{1101}&a_{1111}
                             \end{pmatrix}.
                     \end{equation}
Let $(A_1,A_2,A_3,A_4)$ be an element of the group SLOCC$=SL_2(\CC)^{\times 4}$. The action on $M_\varphi$ is given by $(A_1\otimes A_2) M_\varphi(A_3\otimes A_4)^t$.
This action on the space of $4\times 4$ matrices becomes an action of $SO(4)\times SO(4)$ on $\mathfrak{p}$. This  is described in Ref\cite{CD} via the following unitary matrix (see also Ref \cite{VDMV}):
\begin{equation}
 T=\dfrac{1}{\sqrt{2}}\begin{pmatrix}
    1 & 0 & 0& 1\\
    0 & i & i & 0\\
    0 & -1 & 1 & 0\\
    i & 0 & 0 & -i
   \end{pmatrix}.
\end{equation}
One gets the correspondence: $(A_1,A_2,A_3,A_4)\in \text{SLOCC} \text{ acts on } \mathcal{H}$ and  $\begin{pmatrix}
                                            P_1 & 0\\
                                            0 & P_2
                                           \end{pmatrix}\in SO(4)\times SO(4)$ acts on $\mathfrak{p}$ with $P_1=T(A_1\otimes A_2)T^\dagger$ and $P_2=T(A_3\otimes A_4)T^\dagger$ such that 
                                           \begin{equation}
                                            A_1\otimes A_2\otimes A_3\otimes A_4\ket{\varphi}=\begin{pmatrix}
                                                                                            P_1 & 0 \\
                                                                                            0 & P_2
                                                                                           \end{pmatrix}\begin{pmatrix}
             0 & TM_{\varphi} T^\dagger \\
             -(TM_\varphi T^\dagger)^t& 0
\end{pmatrix}\begin{pmatrix}
P_1 & 0\\
0 & P_2
\end{pmatrix}^{-1}.
                                           \end{equation}

The action of $SO(4)\times SO(4)$ on $\mathfrak{p}$ is nothing but the trace of the adjoint action of $SO(8)$ on  $\mathfrak{p}$.
This embedding of the vector space $\mathcal{H}$ into $\mathfrak{so}_8$ leads to the following observation.
\begin{prop}
 Let $\C[\mathfrak{so}_8]^{SO(8)}$ be the ring of invariant polynomials on $\mathfrak{so}_8$ for the adjoint action of $SO(8)$ and $\CC[\mathcal{H}]^{\text{SLOCC}}$ the ring 
 of invariant polynomials 
 for the SLOCC action on the four-qubit Hilbert space.
 The restriction map \begin{equation}\C[\mathfrak{so}_8]^{SO(8)}\longrightarrow\CC[\mathfrak{p}]^{SO(4)\times SO(4)}=\CC[\mathcal{H}]^{\text{SLOCC}}\end{equation} is an isomorphism.
 Moreover the restriction of the equation defining the dual variety of the adjoint orbit $X_{SO(8)}\subset \PP(\mathfrak{so}_8)$ to $\mathfrak{p}$ is the $2\times 2\times 2\times 2$ hyperdeterminant.
\end{prop}

\proof It is well known\cite{PV} that the ring of invariant polynomials $\CC[\mathfrak{so}_{2n}]^{SO(2n)}$ is a free algebra generated by homogeneous algebraically 
independent polynomials   of degree $2,4,\dots, 2(n-1), n$. Let us denote by $\mathcal{M}^0_{2n}(\CC)$ the space of traceless matrices of size $2n\times 2n$, and let us 
recall that $\mathfrak{so}_{2n}=\{A\in \mathcal{M}^0_{2n}(\CC),A=-A^t\}$. The generators of $\CC[\mathfrak{so}_{2n}]^{SO(2n)}$ can be obtained as the restriction to 
$\mathfrak{so}_{2n}$ of the sum of the $i\times i$ principal minors.
For skew symmetric matrices only the even dimensionnal principal minors do not vanish. Moreover the determinant of a skew symmetric matrix factorizes as the square of the Pfaffian, $Pf$.
In other words the generators of $\CC[\mathfrak{so}_{2n}]^{SO(2n)}$ can be chosen to be $h_2,\dots, h_{2n-2}, Pf$ where $h_{2i}$ is the sum of the $2i\times 2i$ principal minors. 
In particular if 
we consider $\mathfrak{so}_8$, the generators are four polynomials of degree $2,4,6,4$.
The ring of invariant polynomials for four-qubit states under SLOCC\cite{LT} is also a free algebra generated by homogeneous algebraically independent polynomials of degree $2,4,4,6$. Let us denote them respectively 
$B, L, M$ and $D$, their descriptions will be given in the next section. The restriction of the sum of principal minors $h_2, h_4, h_6$ and the Pfaffian $Pf$ to $\mathfrak{p}$, i.e. 
to the vector space \begin{equation}
                                                                                                                                                                      \begin{pmatrix}
                                                                                                                                                                       0 & TM_\varphi T^{\dagger}\\
                                                                                                                                                                       -(TM_\varphi T^{\dagger})^t & 0
                                                                                                                                                                      \end{pmatrix}
                                                                                                                                                                     \end{equation}
         leads to  the following equalities: ${h_2}_{|_\mathfrak{p}}=2B$, ${h_4}_{|_\mathfrak{p}}=B^2+2L+4M$, ${h_6}_{|_\mathfrak{p}}=2BL+4B M-4 D$ and $Pf_{|_\mathfrak{p}}=L$.
 It immediately proves that ${\CC[\mathfrak{so}_8]^{SO(8)}}_{|\mathfrak{p}}=\CC[\mathcal{H}]^{\text{SLOCC}}$.
 
 Regarding the dual variety of $X_{SO(8)}\subset \PP(\mathfrak{so}_8)$, we recall that an equation of $X_{SO(2n)}^*$ is known in terms for the simple roots of the Lie algebras $\mathfrak{so}_{2n}$.
 In fact for all simple Lie algebras the equation defining the dual of the projectivization of the adjoint orbit is given by the vanishing of the discriminant $D_\mathfrak{g}$, i.e. the 
 product of the (long) roots of $\mathfrak{g}$ (see Ref\cite{Tev} p29).
 \begin{equation}
 D_\mathfrak{g}= \prod_{\alpha\in R_l} \alpha=0
 \end{equation}
Let $M\in \mathfrak{so}_8$ and $M_s$ its semi-simple part with eigenvalues $\lambda_1,\dots,\lambda_n$. The roots\cite{F-H} of $\mathfrak{so}_{2n}$ are linear forms on $\mathfrak{h}$, 
the Cartan subalgebra  of $\mathfrak{so}_{2n}$, of the form $\pm L_i\pm L_j$ with $i\neq j$ and such that $L_i(M_s)=\lambda_i$.
Thus if $\alpha$ is a root we have $\alpha(M_s)=\pm\lambda_i\pm \lambda_j$ and the matrix $M\in \mathfrak{so}_{2n}$ belongs
to $X_{SO_{2n}}^*$ if and only if
\begin{equation}
 D_l(M)=\prod_{1\leq i<j\leq n, i\neq j} (\lambda_i\pm\lambda_j)^2=0.
\end{equation}
The roots of the characteristic polynomial of $M$, $t^{2n}+h_2(M)t^{2n-2}+\dots+h_{2n-2}(M)t^2+Pf^2(M)$ are $\pm \lambda_i$. Using the change of variable $x=t^2$, one  gets the polynomial
\begin{equation}\label{firstquartic}
 x^n+h_2(M)x^{n-1}+\dots+h_{2n-2}(M)x+Pf^2(M)
\end{equation}
whose roots are $\lambda_i ^2$. Taking the discriminant one obtains:
\begin{equation}
 \Delta(x^n+h_2(M)x^{n-1}+\dots+h_{2n-2}(M)x+Pf^2(M))=\prod_{1\leq i<j\leq n}(\lambda_i^2-\lambda_j^2)^2=\prod_{1\leq i<j\leq n, i\neq j} (\lambda_i\pm\lambda_j)^2
\end{equation}
One concludes that $\Delta(x^n+h_2(M)x^{n-1}+\dots+h_{2n-2}(M)x+Pf^2(M))=D_{\mathfrak{so}_{2n}}(M)$.

In particular $\Delta(x^4+h_2(M)x^{3}+h_4(M)x^2+h_{6}(M)x+Pf^2(M))=0$ is the defining equation of $X_{SO(8)}^*$. But it can also be checked that 
$\Delta(x^4+2Bx^3+(B^2+2L+4M)x^2+(2BL+4B M-4 D)x+L^2)=0$ is an equation for the dual variety of $\PP^1\times\PP^1\times\PP^1\times\PP^1$, i.e. $\Delta(x^4+2Bx^3+(B^2+2L+4M)x^2+(2BL+4B M-4 D)x+L^2)$ is 
 the hyperdeterminant of format $2\times 2\times 2\times 2$. $\Box$

\begin{rem}\rm
The connection between the SLOCC orbit structure of four-qubits and the Lie algebra $\mathfrak{so}_8$ is also investigated in a different manner in the work of P\'eter L\'evay\cite{Levay1,Levay2} and more recently in Ref\cite{Levay3} with a construction 
of the four-qubit invariants from the spin representation of $SO(16)$. From a different perspective a connection between four-qubit states and the Dynkin diagram $D_4$ (the  Dynkin diagram of $\mathfrak{so}_8$) was also 
pointed out in Ref\cite{HLP} by constructing simple hypersurface singularities of type $D_4$  and their deformations from four-qubit states.
\end{rem}

\begin{rem}\rm
The embedding of $\mathcal{H}$ as a subspace of $\mathfrak{so}_8$ depends on the embedding of the four-qubit state $\ket{\varphi}$ in $\mathcal{M}_4(\CC)$ given by Eq (\ref{embedd}). There are two more ways of looking at a four-qubit state as a linear map from $\CC^4$ to $\CC^4$.
Those other two representations will give different ways of writing the generators of $\CC[\mathcal{H}]^{\text{SLOCC}}$, but also two different types of quartics like the one given by 
Eq (\ref{firstquartic}) for $n=4$. In the next section their role in the classification of the SLOCC orbits is investigated.
\end{rem}

\section{A first classification based on invariants}\label{invariants}
\subsection{SLOCC-invariant polynomials}
In a general setting, a pure $k$-qudit state is an element of the Hilbert space $\mathcal H=V_1\otimes\cdots\otimes V_k$ with $V_i=\mathbb C^{n_i}$ regarded as a multilinear form. 
Two qudit states are equivalent if they belong to the same orbit for the group $\mathrm{SLOCC}=GL_{n_1}\times\cdots\times GL_{n_k}$. In principle, one can determine if two states are 
equivalent by comparing their evaluations on a sufficiently large system of special polynomials (in the coefficients of the forms and auxiliary variables), called concomitants, which are invariant under the action of $\SLOCC$. In practice, this algorithm can be used only in very few cases, because the number and the size of the polynomials increase exponentially with the number of particles and the dimension of the Hilbert space. Nevertheless, even the knowledge of a little part of the polynomials gives rise to interesting information about the classification.

In the case of a four-qubit system, we deal with the quadrilinear form:
\begin{equation}
f:=\sum_{0\leq i,j,k,\ell\leq 1}a_{ijk\ell}x_iy_jz_kt_\ell
\end{equation}
and the algebra of the polynomial invariants (polynomials in the coefficients of the forms with no auxiliary variables) 
is a free algebra on four generators:
\begin{enumerate}
\item One of degree 2: 
\begin{equation}\begin{array}{rcl}B&:=&a_{0000}a_{1111}-a_{1000}a_{0111}+a_{0100}a_{1011}+a_{1100}a_{0011}-a_{0010}a_{1101}
\\&&+a_{1010}a_{0101}-a_{0110}a_{1001}+a_{1110}a_{0001},\end{array}\end{equation}
\item two of degree 4:
\begin{equation}
L:=\left|\begin{array}{cccc}
a_{0000}&a_{0010}&a_{0001}&a_{0011}\\
a_{1000}&a_{1010}&a_{1001}&a_{1011}\\
a_{0100}&a_{0110}&a_{0101}&a_{0111}\\
a_{1100}&a_{1110}&a_{1101}&a_{1111}
\end{array}\right|\mbox{ and } 
M:=\left|\begin{array}{cccc}
a_{0000}&a_{0001}&a_{0100}&a_{0101}\\
a_{1000}&a_{1001}&a_{1100}&a_{1101}\\
a_{0010}&a_{0011}&a_{0110}&a_{0111}\\
a_{1010}&a_{1011}&a_{1110}&a_{1111}
\end{array}\right|
\end{equation}
\item  and one of degree 6: Set $b_{xy}:=\det\left(\dfrac{\partial^2 f}{\partial z_i\partial t_j}\right)$. By interpreting this quadratic form as   a bilinear form on the three dimensional space, one  finds a $3\times 3$ matrix $B_{xy}$ satisfying
$
b_{xy}=[x_0^2,x_0x_1,x_1^2]B_{xy}\left[\begin{array}{c}y_0^2\\y_0y_1\\y_1^2 \end{array}\right].
$
The generator of degree $6$ is \begin{equation}D_{xy}:=-\det(B_{xy}).\end{equation}
\end{enumerate}
Remark that this set is not unique, for instance one can replace $L$ or $M$ by
\begin{equation}N:=-L-M=\left|\begin{array}{cccc}
a_{0000}&a_{1000}&a_{0001}&a_{1001}\\
a_{0100}&a_{1100}&a_{0101}&a_{1101}\\
a_{0010}&a_{1010}&a_{0011}&a_{1011}\\
a_{0110}&a_{1110}&a_{0111}&a_{1111}
\end{array}\right|.
\end{equation}
We remark that, when evaluated on the Verstraete normal form \cite{VDMV}
\begin{equation}\begin{array}{rcl}G_{abcd}&:=&\frac12(a+d)(|0000\rangle+|1111\rangle)+\frac12(a-d)(|1100\rangle+|0011\rangle)+\\&&\frac12(b+c)(|0101\rangle+|1010\rangle)+\frac12(b-c)(|0110\rangle+|0011\rangle,\end{array}\end{equation}  these invariants have nice closed expressions:
\begin{eqnarray}
B(|G_{abcd}\rangle)=\frac12\left(a^2+b^2+c^2+d^2\right),\\
L(|G_{abcd}\rangle)=abcd,\\ M(|G_{abcd}\rangle)=\frac1{16}(a+b+c+d)(c+d-a-b)(a-b+c-d)(a-b+d-c),\\ N(|G_{abcd}\rangle)=\frac1{16}(a+b+c-d)(a-b-c-d)(a-b+c+d)(a+b-c+d),\\
D_{xy}(|G_{abcd}\rangle)=\frac1{32}\, \left( {b}^{2}-{a}^{2}+{c}^{2}-{d}^{2} \right)  \left( -{b}^{
2}+{a}^{2}+{c}^{2}-{d}^{2} \right)  \left( {b}^{2}+{a}^{2}-{c}^{2}-{d}
^{2} \right) 
.
\end{eqnarray}
We consider the space $\mathcal{S}:=\left\{\left|G_{abcd}\right\rangle:(a,b,c,d)\in\C^4\right\}$ of the Verstraete normal forms.  It is the Chevalley section for  the $\SLOCC$ action on  the Hilbert space 
with Weyl group $D_4$: the closure of each generic orbit intersects $\mathcal{S}$ along a $D_4$ orbit. 
Indeed, since the polynomials $B(|G_{abcd}\rangle)$, $L(|G_{abcd}\rangle)$, $M(|G_{abcd}\rangle)$, and $D_{xy}(|G_{abcd}\rangle)$ are algebraically independent, the system 
\begin{equation}
\left\{
\begin{array}{l}
B(|G_{abcd}\rangle)=\alpha\\
L(|G_{abcd}\rangle)=\beta\\
M(|G_{abcd}\rangle)=\gamma\\
D_{xy}(|G_{abcd}\rangle)=\delta
\end{array}
\right.
\end{equation}
 admits at most $192$ solutions (this is the order of $D_4$). Furthermore, the system is clearly invariant under the permutations of the variables $(a,b,c,d)$ and the transformation $(a,b)\longrightarrow (-a,-b)$. For generic values of $(a,b,c,d)$, these transformations generate a group isomorphic to $D_4$. Hence, knowing one solution, the other  ones are deduced from the action of $D_4$. Geometrically, the solutions are symmetric with respect to the reflection group of the demitesseract.
See Appendix \ref{demitesseract}. 


\subsection{Three quadrics}

We consider the three quartics
\begin{eqnarray}
Q_1(|\varphi\rangle):=x^4-2Bx^3y+(B^2+2L+4M)x^2y^2+(4D_{xy}-4B(M+\frac12L))xy^3+L^2y^4,\\
Q_2(|\varphi\rangle):=x^4-2Bx^3y+(B^2-4L-2M)x^2y^2+(-2MB+4D_{xy})xy^3+M^2y^4,\\
Q_3(|\varphi\rangle):=x^4-2Bx^3y+(B^2+2L-2M)x^2y^2-(2LB+2MB-4D_{xy})xy^3+N^2y^4.
\end{eqnarray}
Evaluated on $|G_{abcd}\rangle$, the roots of $Q_1$  are  $a^2$, $b^2$, $c^2$ and $d^2$ and the roots of $Q_2$ (resp. $Q_3$) are the squares of the four  polynomial factors of $M(|G_{abcd}\rangle)$ (resp. $N(|G_{abcd}\rangle)$ which are obtained by applying an invertible linear transformation to $(a,b,c,d)$. Hence, the three quartics have the same invariants.\\
The invariant polynomials of a quartic $f:=\alpha x^4-4\beta x^3y+6\gamma x^2y^2-4\delta xy^3+\omega y^4$ are algebraic combinations of $I_2=\alpha\omega-4\beta\delta+3\gamma^2$, an invariant of degree $2$, which is the apolar of the form with itself, and $I_3=\alpha\gamma\omega-\alpha\delta^2-\beta^2\omega-\gamma^3+2\beta\gamma\omega$, an invariant of degree $3$ called 
the catalecticant\cite{Olver}.
In particular, the discriminant of the quadric is $\Delta=I_2^3-27I_3^2$.\\
From these definitions, one has
\begin{equation}
I_2(Q_1)=I_2(Q_2)=I_2(Q_3)=
\frac43\,{{L}}^{2}+2\,{B}\,{D_{xy}}-\frac43\,{{B}}^{2}{M}-\frac23\,{{B}}^{2}{L}+\frac1{12}\,{{B}}^{4}+\frac43\,{L}\,{
M}+\frac43\,{{M}}^{2},
\end{equation}
and
\begin{equation}
\begin{array}{rcl}
I_3(Q_1)=I_3(Q_2)=I_3(Q_3)&=&
\frac42\,{D_{xy}}\,{B}\,{M}+\frac23\,{D_{xy}}\,{B}\,{L
}-\frac59\,{{B}}^{2}{M}\,{L}+\frac49\,{L}^{2}{M}-{{
D_{xy}}}^{2}-\frac59\,{{B}}^{2}{{M}}^{2}\\&&+\frac1{18}\,{B}^{4}{
L}+\frac19\,{B}^{4}{M}-\frac49\,{L}\,{{M}}^{2}-\frac16
\,{{B}}^{3}{D_{xy}}-\frac29\,{{B}}^{2}{{L}}^{2}+{\frac {8
}{27}}\,{{L}}^{3}\\&&-{\frac {1}{216}}\,{{B}}^{6}-{\frac {8}{27}
}\,{{M}}^{3}.
\end{array}
\end{equation}
Furthermore, we easily check that $\Delta(Q_i)$ is also the hyperdeterminant of $|\varphi\rangle$ (regarded as a quadrilinar form).
\\
In the aim to describe the roots of the quartics, we will use two other covariant polynomials: the Hessian
\begin{equation}
Hess(f):=\left|\begin{array}{cc}{\partial^2\over \partial x^2}f&{\partial^2\over \partial x\partial y}f\\
{\partial^2\over \partial y\partial x}f&{\partial^2\over \partial y^2}f\end{array}\right|
\end{equation}
and the Jacobian of the Hessian
\begin{equation}
T(f)=\left|\begin{array}{cc}{\partial\over \partial x}f&{\partial\over \partial y}f\\
{\partial\over \partial x}Hess(f)&{\partial\over \partial y}Hess(f)\end{array}\right|.
\end{equation}
From the values of the covariants one can compute the multiplicity of the roots of a quartic $f$, according to  Table \ref{RootQuart}.
\begin{table}\[
\begin{array}{|c|l|}
\hline 
Covariants&Interpretation\\
\hline
\Delta\neq 0&\mbox{ Four distinct roots}\\
\Delta=0\mbox{ and }T\neq 0&\mbox{ Exactly one double root}\\
T=0\mbox{ and }I_2\neq 0&\mbox{ Two distinct double roots}\\
I_2=I_3=0\mbox{ and }Hess\neq 0&\mbox{ A triple root}\\
Hess=0&\mbox{ A quadruple root}\\\hline
\end{array}\]
\caption{Roots of a quartic\label{RootQuart}}
\end{table}
Notice that the evaluations of $Hess$ and $T$ on the forms $Q_1$, $Q_2$ and $Q_3$ are not equal in the general case.
\subsection{A first classification}
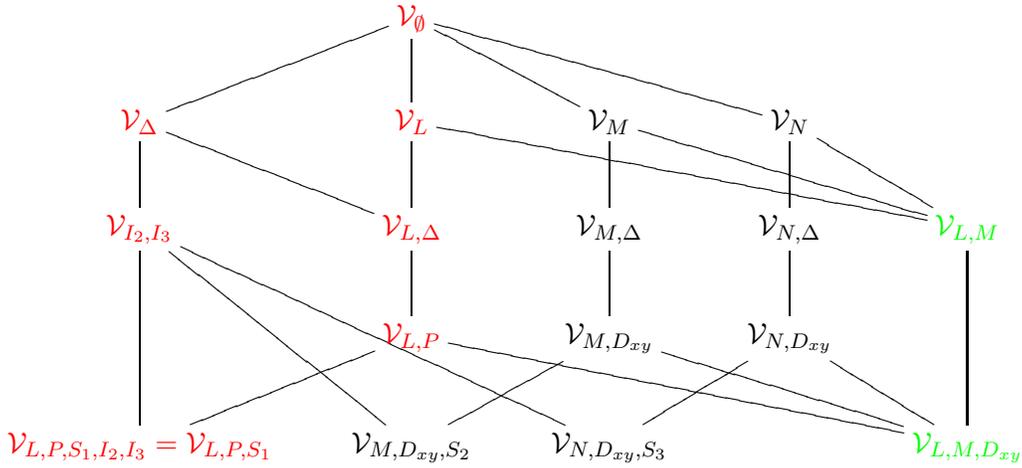
\begin{figure}[!h]
  \[\xymatrix{&\color{red}\mathcal V_\emptyset\incl@[red][dl]\incl@[red][d]\incl[dr]\incl[drr]\\
 \color{red}\mathcal V_\Delta\incl@[red][d]\incl@[red][dr]&\color{red}\mathcal V_L\incl@[red][d]\incl[drrr]&\mathcal V_M\incl[d]\incl[drr]&\mathcal V_N\incl[d]\incl[dr]\\
\color{red}\mathcal V_{I_2,I_3}\incl@[red][dd]\incl[ddr]\incl[ddrr]&\color{red}\mathcal V_{L,\Delta}\incl@[red][d]&\mathcal V_{M,\Delta}\incl[d]&\mathcal V_{N,\Delta}\incl[d]&\color{green}\mathcal V_{L,M}\incl@[green][dd]\\
&\color{red}\mathcal V_{L,P}\incl@[red][dl]\incl[drrr]&\mathcal V_{M,D_{xy}}\incl[dl]\incl[drr]&\mathcal V_{N,D_{xy}}\incl[dl]\incl[dr]\\
\color{red}\mathcal V_{L,P,S_1,I_2,I_3}=\mathcal V_{L,P,S_1}&\mathcal V_{M,D_{xy},S_2}&\mathcal V_{N,D_{xy},S_3}&&\color{green}\mathcal V_{L,M,D_{xy}}
}\]
\caption{Inclusion diagram of the varieties $\mathcal V_S$.}\label{VDiag}
\end{figure}

\begin{figure}[!h]
\[ \xymatrix{&\mbox{Generic}\incl[dl]\incl[d]\\
 \mbox{a double root}\incl[d]\incl[dr]&Q_1(0,y)=0\incl[d]\\
\mbox{a triple root}\incl[dd]&{\mbox{a double root}\atop Q_1(0,y)=0}\incl[d]\\
&Q_1\mbox{ has a double zero root}\incl[dl]\incl[dr]\\
Q_1\mbox{ has a zero triple root}&&\color{green}Q_1\mbox{ has a double zero root}\atop Q_2\mbox{ has a zero root}
}\]
\caption{Interpretation of the red part of Figure \ref{VDiag} in terms of quartics.\label{VDiagQuart}}
\end{figure}
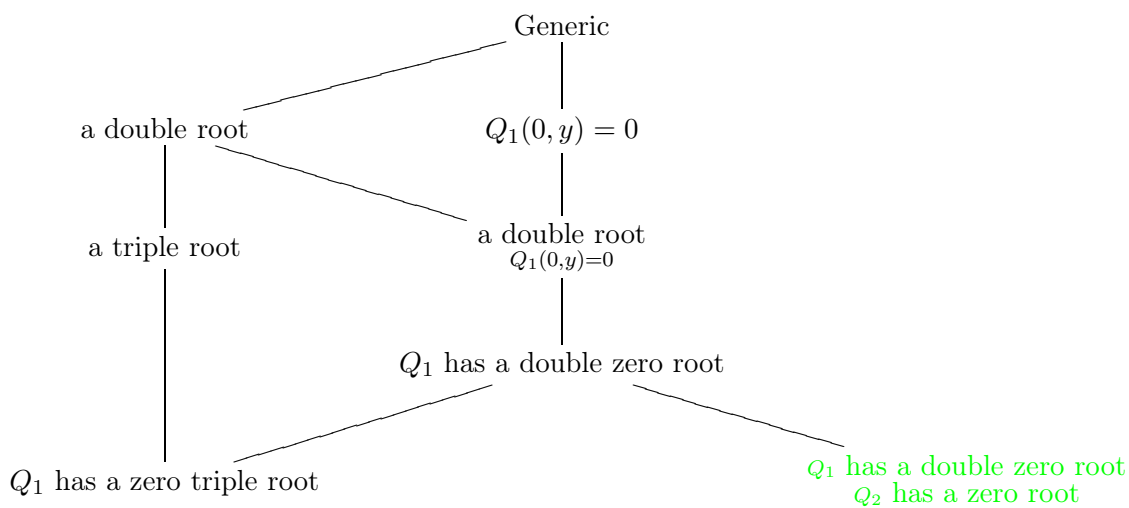

In this section, we use Table \ref{RootQuart}  to obtain a first classification and we refine it by considering the polynomials $L$, $M$ and $N$ 
which allow to decide if a quartic has a null root. The discussion is relegated to Appendix \ref{DiscRoots}.\\
We define the invariants 
\begin{equation}
P:=D_{xy}-BM, S_1:=B^2+4M, S_2=B^2-4L,\mbox{ and }S_3=B^2-4M.
\end{equation}
If $S$ is a set of polynomials, we will denote by $\mathcal V_S$ the variety defined by the system $\{E=0:E\in S\}$. In a previous paper\cite{HLT2}, we 
have investigated the case when $L=M=0$. It remains to consider the other cases and, according to Appendix \ref{DiscRoots}, one has to refine the diagram of inclusions  
of Figure \ref{VDiag} which represents a first tentative of classification (see also Figure \ref{VDiagQuart} for the 
interpretation of red part in terms of roots of the quartics). The green part corresponds to subvarieties of $\mathcal V_{L,M}$ 
that are already  investigated in a previous paper\cite{HLT2}. Also, notice that the whole diagram of Figure \ref{VDiag} can be
deduced from the red part,  replacing $L=0$ by $M=0$  (resp. $N=0$). Note  that $\mathcal V_{L,P,S_1,I_2,I_3}=\mathcal V_{L,P,S_1}$. Indeed, for any form in $\mathcal V_{L,P,S_1,I_2,I_3}$, $Q_1$ has a zero triple root and this implies automatically that $Q_2$ has a quadruple root which is equal to one of the parameter of the normal 
form. The case where one of the quartic has two double roots does not appear explicitly in the diagrams. But, a short calculation shows that it is equivalent to the case where one of the quartics has zero as a double root.
\subsection{The form problem again}
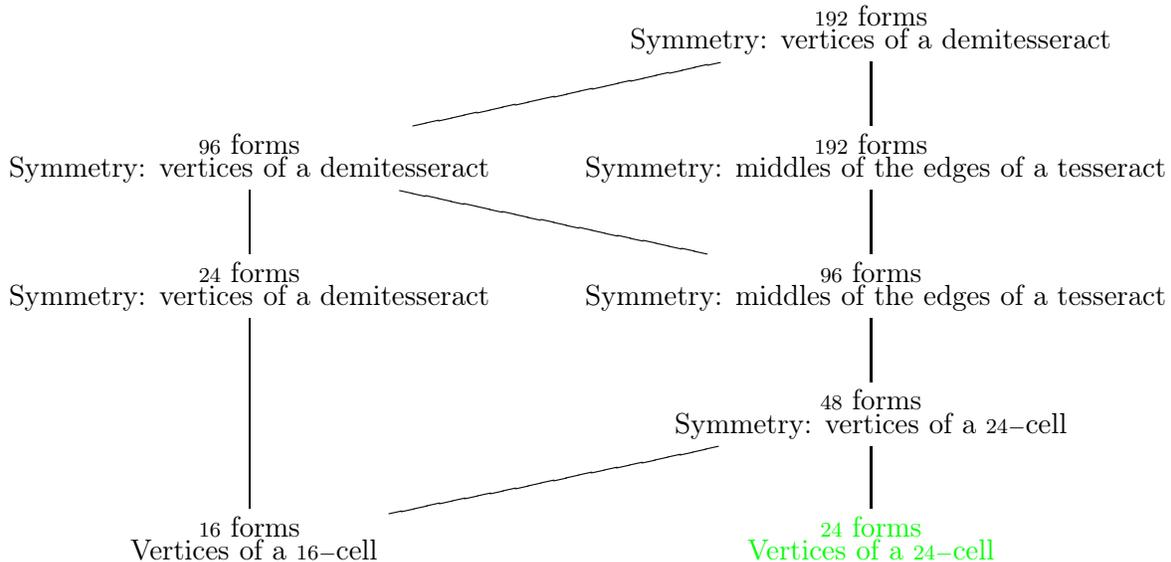
\begin{figure}[!h]
  \[\xymatrix{&{192\mbox{ forms}\atop\mbox{Symmetry: vertices of a demitesseract}}\incl[dl]\incl[d]\\
 {96\mbox{ forms}\atop\mbox{Symmetry: vertices of a demitesseract}}\incl[d]\incl[dr]&192\mbox{ forms}\atop\mbox{ Symmetry: middles of the edges of a tesseract}\incl[d]\\
 {24\mbox{ forms}\atop\mbox{Symmetry: vertices of a demitesseract}}\incl[dd]&96\mbox{ forms}\atop\mbox{ Symmetry: middles of the edges of a tesseract}\incl[d]\\
&48\mbox{ forms}\atop\mbox{Symmetry: vertices of a }24-\mbox{cell}\incl[dl]\incl[d]\\
16\mbox{ forms}\atop\mbox{ Vertices of a }16-\mbox{cell}&\color{green}24\mbox{ forms}\atop \color{green} \mbox{Vertices of a } 24-\mbox{cell}
}\]
\caption{Diagram of normal forms.\label{VDiagNF}}
\end{figure}

Each of the varieties described in Figure \ref{VDiag} contains orbits whose intersections with the Chevalley section $\mathcal S$ are finite sets of points with symmetries related to some four dimensional polytopes. 
\begin{itemize}
\item A generic orbit in $\mathcal V_\emptyset$ intersect $\mathcal S$ in $192$ points splitting into $8$ subsets of $24$ points belonging to the same hyperplane. Each of these subsets is constituted with the permutations of the same vector $(a,b,c,d)$ and centered on one of the vertices of a demitesseract (see Figure \ref{generic}).
\begin{figure}[!h]
\begin{center}
\includegraphics[width=5cm]{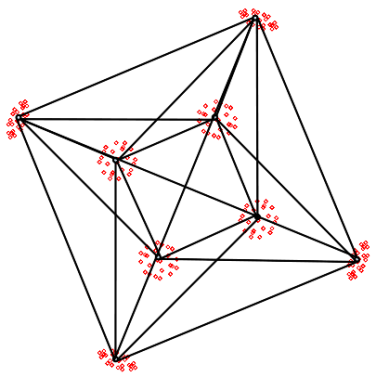} 
\end{center}
\caption{Intersection of a generic orbit of $\mathcal V_\emptyset$ with the subspace of normal forms.\label{generic}}
\end{figure}
  \item In $\mathcal V_L$, each generic orbit contains also $192$ normal forms which are the permutations of the same vector $(0,a,b,c)$. The set of the normal forms splits into $32$ subsets which are constituted of $6$ points, belonging to the same space of dimension $2$, centered on one of the points $(\alpha,\alpha,\alpha,0)$, $(\alpha,\alpha,0,\alpha)$, $(\alpha,0,\alpha,\alpha)$ or $(0,\alpha,\alpha,\alpha)$ with $\alpha={\pm a\pm b\pm c\over 3}$, that are middles of the edges of a tesseract (see Figure \ref{a=0}).
\begin{figure}[!h]
\begin{center}
\includegraphics[width=5cm]{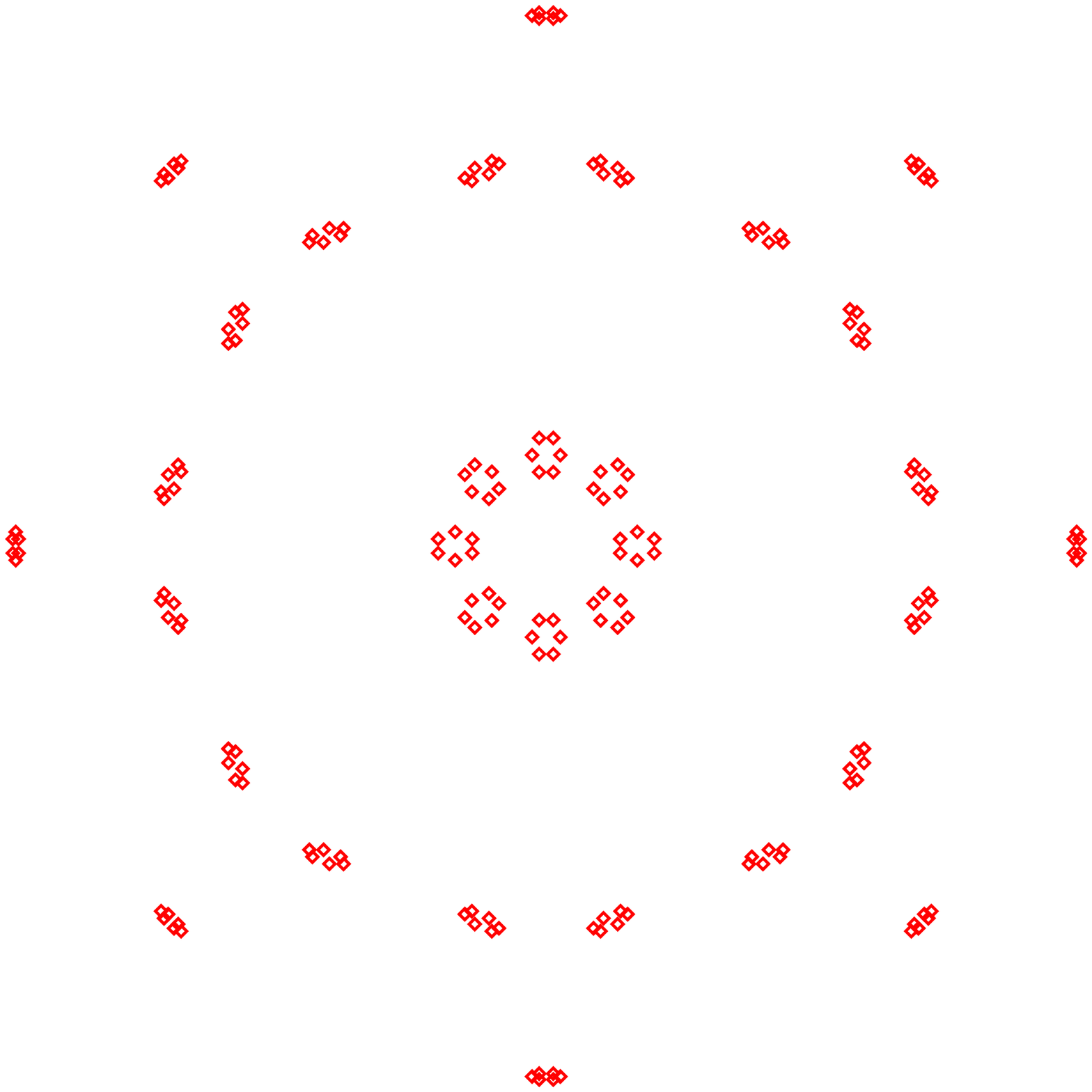} \includegraphics[width=5cm]{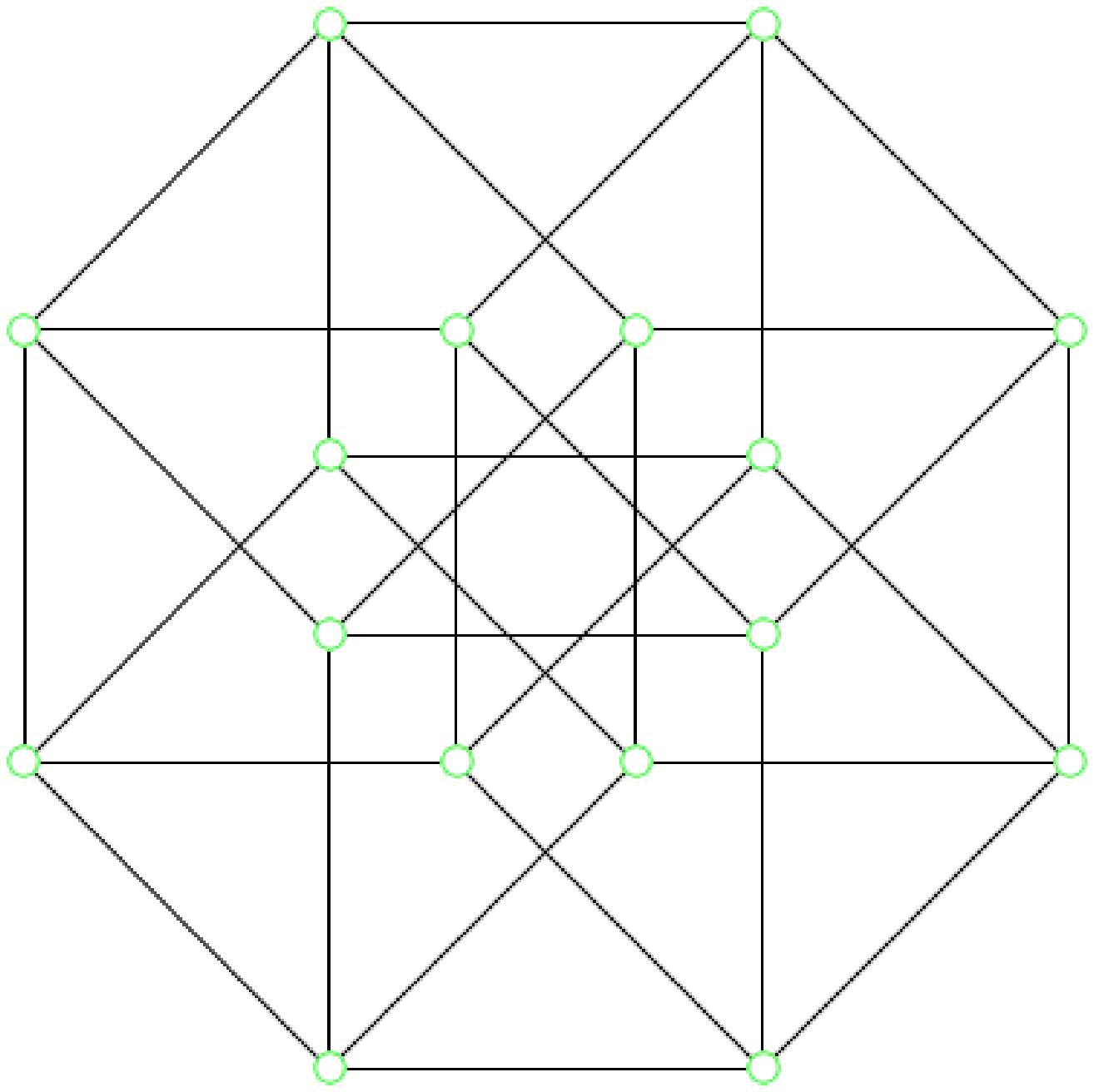} 
\end{center}
\caption{Intersection of a generic orbit of $\mathcal V_L$ with the subspace of normal forms and the tesseract.\label{a=0}}
\end{figure}
\item A generic orbit in $\mathcal V_{\Delta}$ is a special orbit in $\mathcal V_\emptyset$ with $96$ normal forms splitting into $8$ subsets of $12$ points.\\
\item A generic orbit in $\mathcal V_{\Delta,L}$ is a special orbit in $\mathcal V_L$ with $96$ normal forms splitting into $32$ subsets of $3$ points centered on the middle of the edges of a tesseract (see Figure \ref{a=0,b=c}).\\
\begin{figure}[!h]
\begin{center}
\includegraphics[width=5cm]{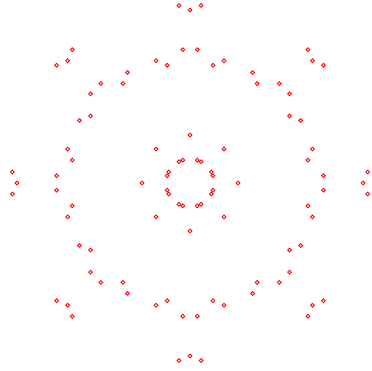}  
\end{center}
\caption{Intersection of a generic orbit of $\mathcal V_{\Delta,L}$ with the subspace of normal forms.\label{a=0,b=c}}
\end{figure}
\item A generic orbit in $\mathcal V_{I_2,I_3}$ is a special orbit in $\mathcal V_{\Delta}$ with $24$ normal forms splitting into $8$ subsets of $4$ points.
\item A generic orbit in  $\mathcal V_{L,P}$ is a special orbit in $\mathcal V_{L,\Delta}$ with $48$ normal forms splitting into $24$ pairs of points whose the middles are 
the vertices of a $24$-cell polytope (see Figure \ref{a=b=0}). These vertices are also the centers of the faces of a tesseract. Notice that such an orbit can degenerate and have only $24$ normal forms, which are exactly the vertices of the $24$-cell. In this case, one has $T(Q_1)=0$ (because $Q_1$ has two double roots). From Appendix \ref{DiscRoots}, it follows that $L=M=0$ and, since this case has already been investigated in one of our previous papers\cite{}, we will not examine it here. 
\begin{figure}[!h]
\begin{center}
\includegraphics[width=5cm]{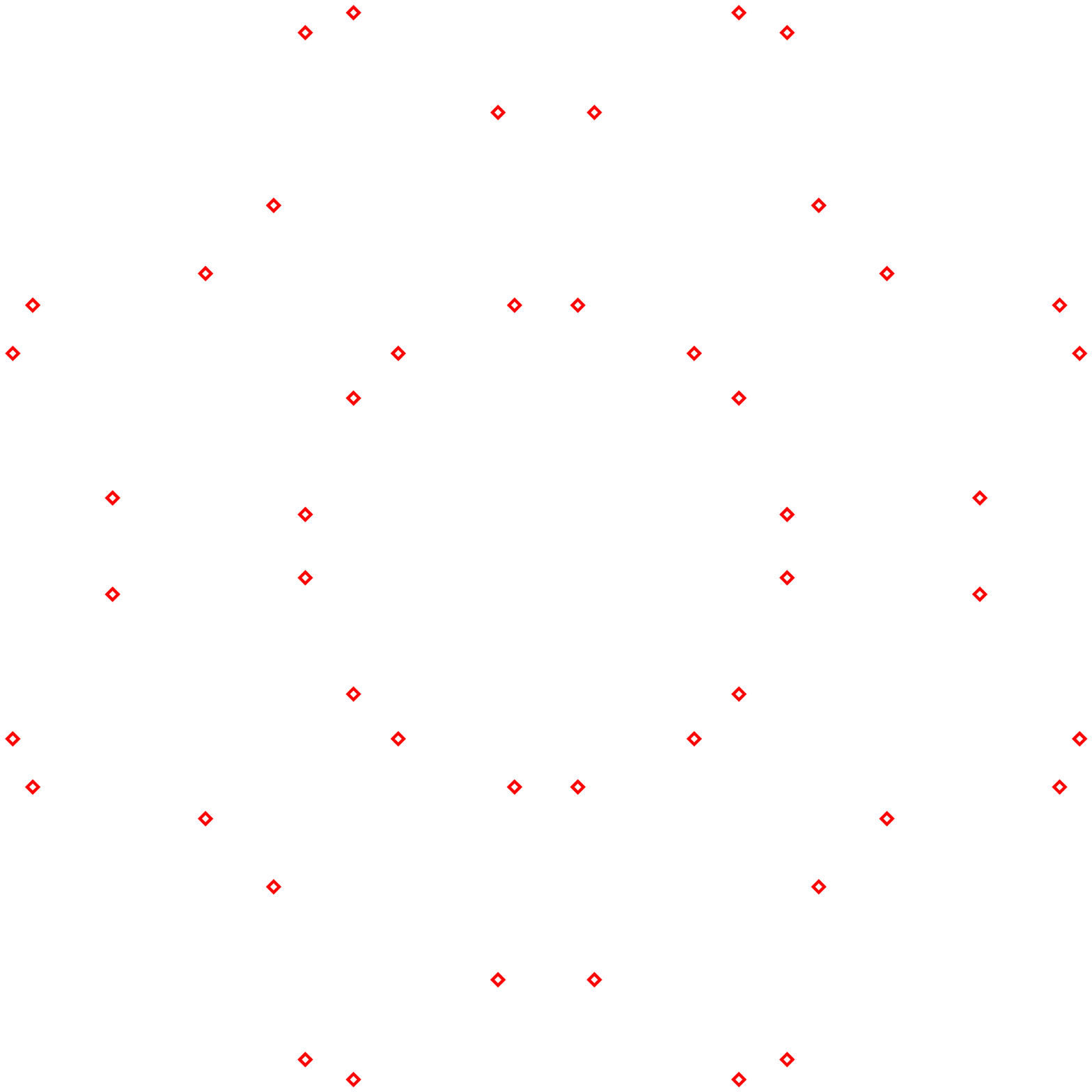}  \includegraphics[width=5cm]{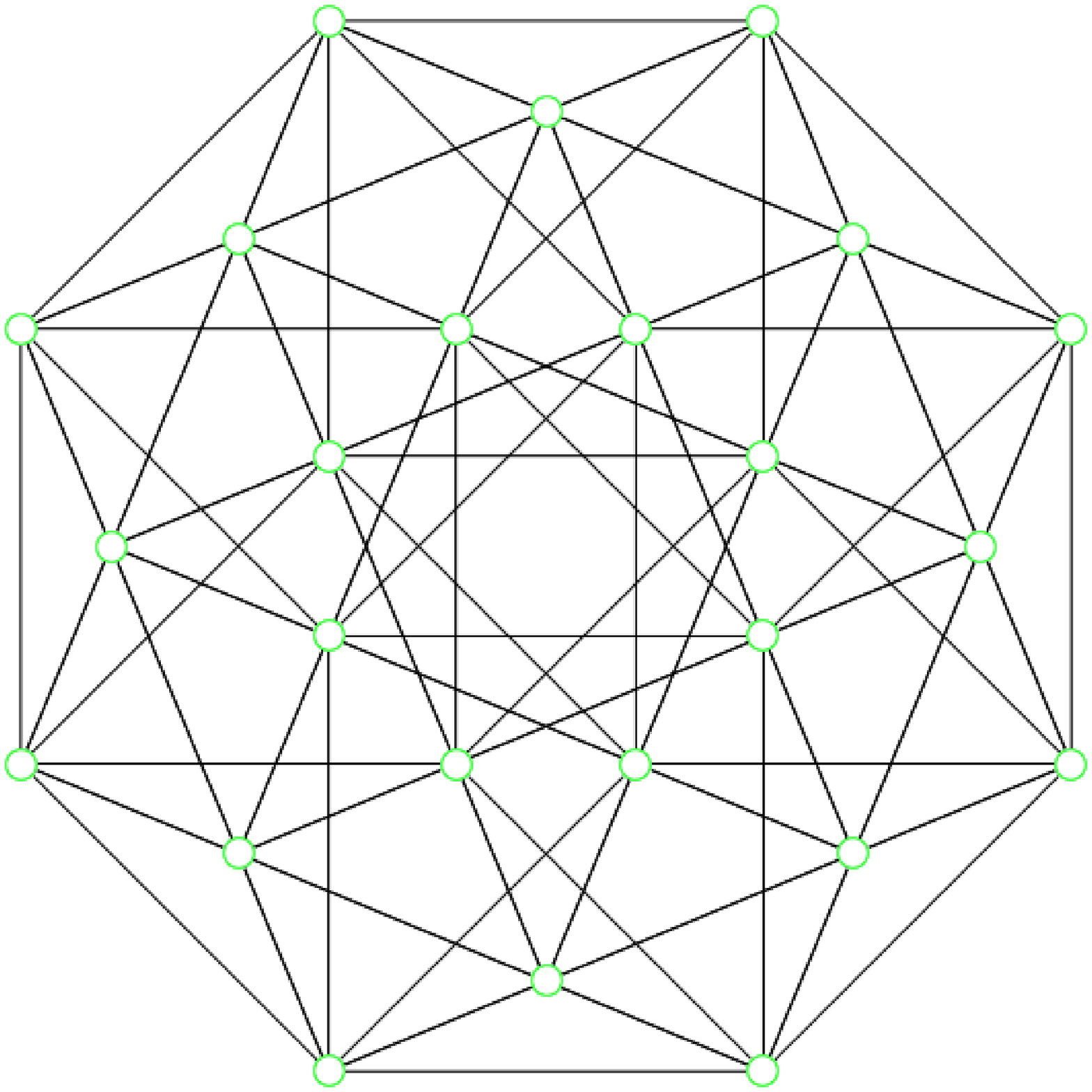}
\end{center}
\caption{Intersection of a generic orbit of $\mathcal V_{L,P}$ with the subspace of normal forms and a $24$-cell polytope.\label{a=b=0}}
\end{figure}
\item  A generic orbit in  $\mathcal V_{L,P,S_1}$ is a special orbit in $\mathcal V_{L,P}$ with $16$ normal forms, which are the vertices of a $16$-cell polytope (see Figure \ref{a=b=c=0}). 
\begin{figure}[!h]
\begin{center}
\includegraphics[width=5cm]{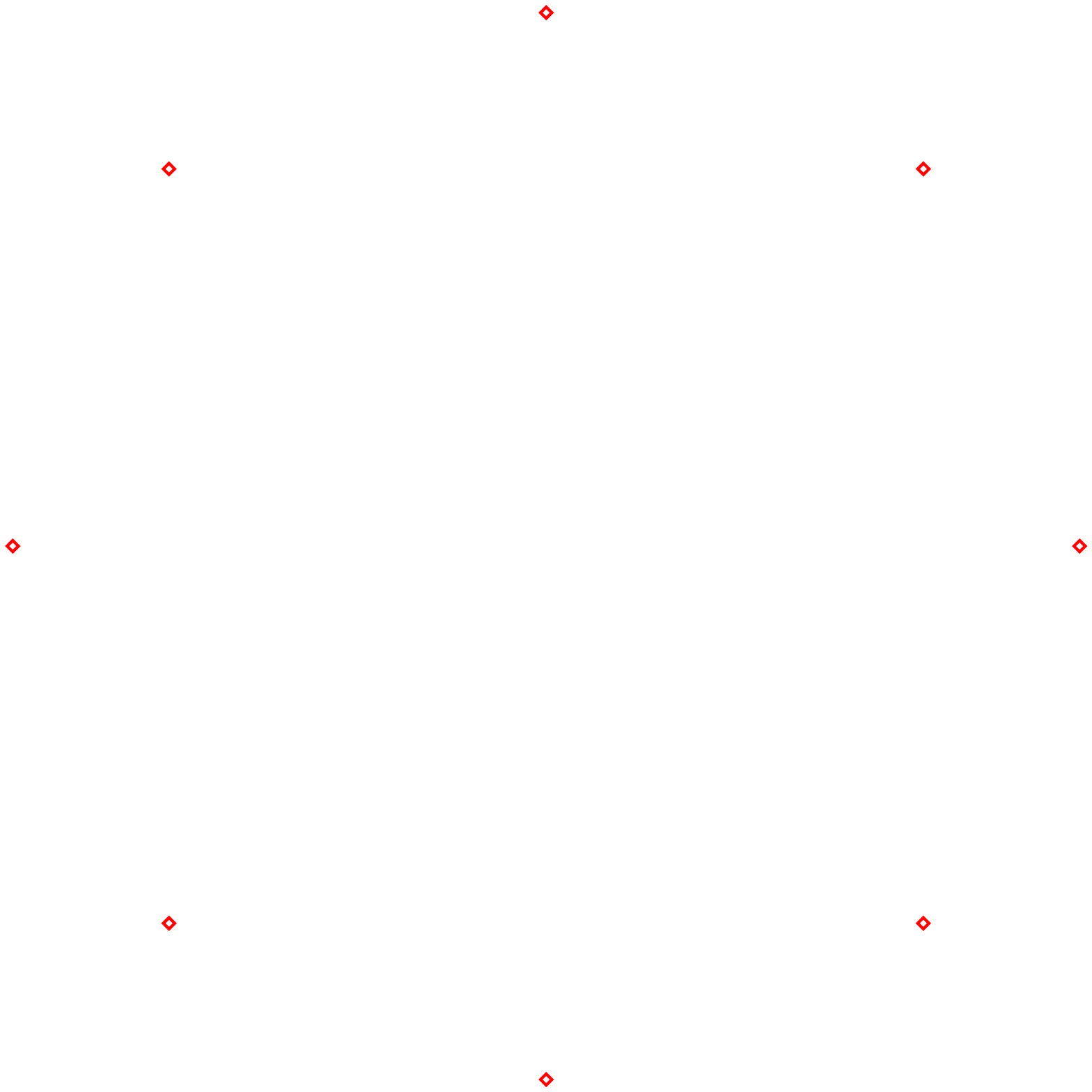}  \includegraphics[width=5cm]{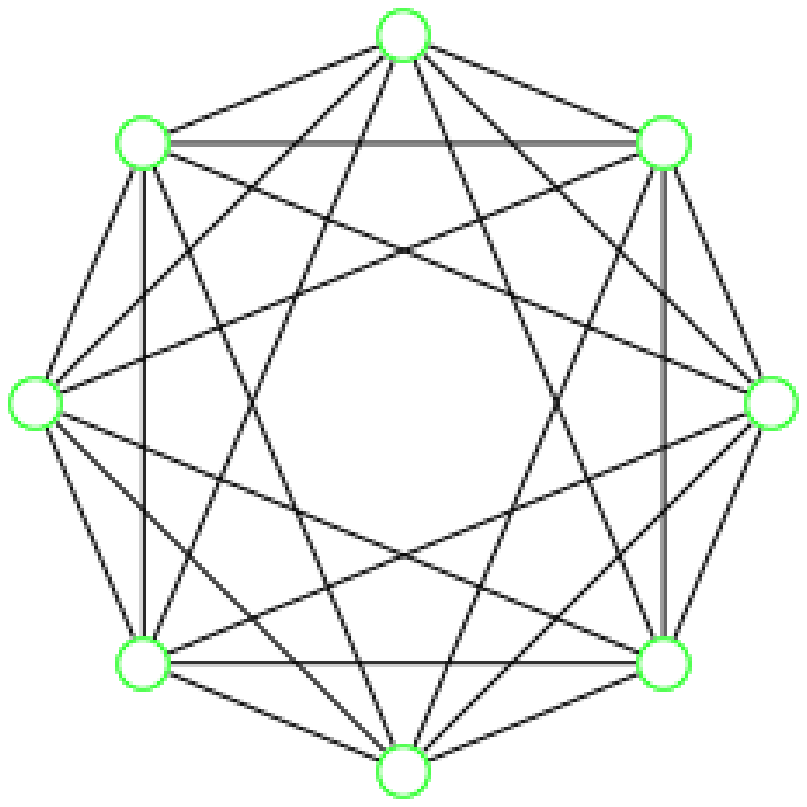}
\end{center}
\caption{Intersection of a generic orbit of $\mathcal V_{L,P,S_1}$ with the subspace of normal forms and a $16$-cell polytope.\label{a=b=c=0}}
\end{figure}
\end{itemize}
Hence, one can interpret the diagram of Figure \ref{VDiag} in terms of normal forms in Figure \ref{VDiagNF}.
Let us end the discussion by an illustration of the fact that the quartics $Q_1$, $Q_2$ and $Q_3$ have an interchangeable role: we first remark that one can choose a degenerate orbit in $\mathcal V_\emptyset$ by setting $a=b=c=d$. In this case, the normal forms are the vertices of the demitesseract and so of a $16$-cell. It is  similar (up to a rotation) to a generic orbit in $\mathcal V_{L,\Delta,P,Q}$. In terms of quartics, this is interpreted by the fact that $Q_1$ has a quadruple root and $Q_2$ has a zero triple root (compare to the variety $\mathcal V_{L,\Delta,P,Q}$, where $Q_1$ has a zero triple root and $Q_2$ has a quadruple root).

\section{Classification based on geometric stratification}\label{geo}
We now investigate the SLOCC structure of the four-qubit Hilbert space from a geometrical point of view. Because a quantum state is well-defined up to a phase factor we will consider the 
projective Hilbert space $\PP(\mathcal{H})=\PP(\CC^2\otimes\CC^2\otimes\CC^2\otimes \CC^2)$ and we recall that $X=\PP^1\times\PP^1\times\PP^1\times\PP^1$, the Segre embedding of 
four projective lines, is the variety 
of separable states\cite{HLT}. We begin the section by defining the auxiliary varieties already used in our previous work\cite{HLT, HLT2} to describe SLOCC invariant varieties built from the knowledge of $X$.
Then we look at various stratifications of the ambient space induced by the singular locus of the four hypersurfaces corresponding to the  zero locus of $L, M, N$ and $\Delta$. When combined, we recover the stratification 
induced by the three quartics $Q_1, Q_2$ and $Q_3$.


\subsection{Secant, tangent and dual varieties}
The basic geometric tool introduced in our previous papers\cite{HLT, HLT2} was the concept of auxiliary varieties. An auxiliary variety is an algebraic variety built by elementary 
geometric constructions 
from a  given variety. An example of such is the so called secant variety\cite{Z2,Lan3}. If $Y\subset \PP(V)$ is a projective variety, the secant variety of $Y$ is the algebraic closure of the union of secant lines:
\begin{equation}
 \sigma(Y)=\overline{\bigcup_{x,y\in Y} \PP^1 _{xy}}.
\end{equation}

If $Y=X$, the set of separable states, then $\sigma(X)$ is the algebraic closure of quantum states which are sums of two separable states. This algebraic variety is SLOCC invariant because so is $X$
and it can be shown, for any multipartite system, that $\sigma(X)=\overline{\text{SLOCC}\ket{GHZ}}$ where $\ket{GHZ}$ is the usual generalization of the GHZ-state and $X$ still denotes the variety of separable states.

In the same spirit, higher dimensional secant varieties can be defined by
\begin{equation}
 \sigma_k(Y)=\overline{\bigcup_{x_1,\dots,x_k\in Y}\PP^{k-1} _{x_1,\dots,x_k}}.
\end{equation}

The secant varieties naturally provides a stratification of the ambient space and states belonging to different secant varieties can not be SLOCC equivalent. 
The interest of secant varieties in the context  of quantum entanglement was first pointed out by Heydari\cite{hey}.

Another auxiliary variety of importance is the so-called tangential variety which corresponds to the union of tangent lines. If $Y$ is a smooth variety and $\tilde{T}_y Y$ denotes the embedded 
tangent space of $Y$ at $y$ we have.
\begin{equation}
 \tau(Y)=\bigcup_{y\in Y} \tilde{T}_y Y.
\end{equation}

This variety has also a nice quantum information theory interpretation. For any multipartite system we have, if $X$ still denotes the variety of separable states, $\tau(X)=\overline{\text{SLOCC}\ket{W}}$ where $\ket{W}$ is 
the generalization of the W-states. Those are well-known fact from algebraic geometry\cite{Z2,Lan3} and have been restated in the language of quantum information theory in Ref\cite{HL}.

In our geometric description of the stratification of the ambient space by algebraic varieties we will also use the concept of dual variety. Let us remind what the dual variety of a projective algebraic variety is, and why 
this concept has been already introduced in the study of entanglement of multipartite systems\cite{My,HLT}.

Consider $X\subset \PP(V)$ a nondegenerate projective variety (i.e. not contained in a hyperplane). The dual variety of $X$ is the closure of the set of tangent hyperplanes, i.e. is defined by 
\begin{equation}
 X^*=\overline{\{H\in \PP(V^*),\exists x\in X_{\text{smooth}}, \tilde{T}_x X\subset H\}}
\end{equation}

where $\tilde{T}_x X$ denote the embedded tangent space of $X$ at $x$ (see Ref\cite{Lan}).

If $X$ is a $G$-invariant variety for a $G$-action on $V$ and $X^*$ is a hypersurface, then the defining equation of $X^*$ is a $G$-invariant polynomial. 
In the case of $X=\PP^{k_1}\times\dots\times \PP^{k_r}$ with 
$k_1\leq k_2+\dots+k_r$ (assuming $k_1\geq k_i$) the dual variety is always a hypersurface called the hyperdeterminant of format 
$(k_1+1)\times (k_2+1)\times\dots\times (k_r+1)$ and it is a $G=GL_{k_1+1}\times\dots\times GL_{k_r+1}$-invariant polynomial.
Hyperdeterminants have been deeply studied by Gelfand Kapranov and Zelevinsky\cite{GKZ1,GKZ}.

In the case where $X=\PP^1\times\PP^1\times\PP^1$, the hyperdeterminant of format $2\times 2\times 2$
 is the so-called Cayley hyperdeterminant and when $X=\PP^1\times\PP^1\times\PP^1\times\PP^1$, the hyperdeterminant is the invariant polynomial $\Delta$ introduced in 
Section \ref{invariants}.

Other SLOCC invariant polynomials, or SLOCC invariant algebraic varieties of the Hilbert space of four qubits can be interpreted in terms of dual varieties as we now show.
\subsection{The hypersufaces $\{L=0\}$, $\{M=0\}$ and $\{N=0\}$}
The three hypersurfaces defined by the vanishing of one of the quartic invariant polynomials $L$, $M$ or $N$ are isomorphic 
and correspond to the dual varieties of three different types of
embeddings of $\PP^3\times \PP^3$ in $\PP^{15}=\PP(\CC^2\otimes\CC^2\otimes\CC^2\otimes \CC^2)$.

Indeed, let $A, B\in \CC^2\otimes\CC^2$ be two $2\times 2$ matrices which are tensors of rank at most two, i.e. $A=t_1\otimes u_1+t_2\otimes u_2$ and $B= v_1\otimes w_1+ v_2\otimes w_2$.
Then we have the following three embeddings:
\[\begin{array}{cccc}
& \PP^3\times\PP^3 & \hookrightarrow & \PP^{15}\\
\phi_1: &([A],[B]) & \hookrightarrow & [ t_1\otimes u_1\otimes v_1\otimes w_1+ t_1\otimes u_1\otimes v_2\otimes w_2\\
& &&+t_2\otimes u_2\otimes v_1\otimes w_1 + t_2\otimes u_2\otimes v_2\otimes w_2]\\
\phi_2: &([A],[B]) & \hookrightarrow & [ w_1\otimes u_1\otimes v_1\otimes t_1+ w_1\otimes u_1\otimes v_2\otimes t_2\\
& &&+w_2\otimes u_2\otimes v_1\otimes t_1 + w_2\otimes u_2\otimes v_2\otimes t_2]\\
\phi_3: &([A],[B]) & \hookrightarrow & [ t_1\otimes v_1\otimes u_1\otimes w_1+ t_1\otimes v_1\otimes u_2\otimes w_2\\
& &&+t_2\otimes v_2\otimes u_1\otimes w_1 + t_2\otimes v_2\otimes u_2\otimes w_2]\\
\end{array}\]
\begin{rem}\rm\rm
There are no other $\phi_i$ to consider. Indeed, more permutations will not give any new varieties. 
For instance exchanging the vectors $t$ and $u$ in $\phi_1$ to construct an other map $\tilde{\phi}_1$ will not give
 anything  new because $\phi_1([A],[B])=\tilde{\phi}_1([^tA],[B])$.
\end{rem}

%
%
%

We will denote by $Seg_i(\PP^3\times\PP^3)\subset\PP(\mathcal{H})$, for $i=1,2,3$, the three Segre embeddings of $\PP^3\times\PP^3$. 
From a QIT perspective, it should be pointed out that the Segre embedding of $\PP^3\times\PP^3\subset \PP^{15}$ corresponds to the algebraic closure of the product 
of two $|EPR\rangle$ states, i.e. for $\phi_1$ we have
\begin{equation}
 \PP^3\times\PP^3=\overline{\text{SLOCC}(|00\rangle+|11\rangle)\otimes (|00\rangle+|11\rangle)}=\overline{\text{SLOCC}(|EPR\rangle\otimes|EPR\rangle)}\subset \PP^{15}.
\end{equation}

In other words the Segre embedding of $\PP^3\times\PP^3$ by $\phi_1$ corresponds to the orbit closure of 
\begin{equation}
 |\varphi_1\rangle=|0000\rangle+|0011\rangle+|1100\rangle+|1111\rangle.
\end{equation}

Similarly, the embeddings provided by $\phi_2$ and $\phi_3$ are the orbit closures of $|\varphi_2\rangle$ and $|\varphi_3\rangle$, with

\begin{equation}
 |\varphi_2\rangle=|0000\rangle+|0101\rangle+|1010\rangle+|1111\rangle 
\end{equation}
and
\begin{equation}
  |\varphi_3\rangle=|0000\rangle+|0110\rangle+|1001\rangle+|1111\rangle.
\end{equation}
Let us denote by $\ket{\varphi^\sigma}$ the state obtained from $\ket{\varphi}$ by permuting the qubits by $\sigma$. Then, it is clear that $\ket{\varphi_2}=\ket{\varphi^{1324}}$ 
and $\ket{\varphi_3}=\ket{\varphi^{1432}}$.
It is well known that the ``usual`` Segre product of two projective spaces $\PP^m\times\PP^n\subset\PP(\CC^{m+1}\otimes\CC^{n+1})$ corresponds to the projectivization of the variety of 
rank one matrices in the projectivization of the space of $(m+1)\times(n+1)$ matrices. Taking the sum of two matrices of rank one, one gets a
 matrix of rank at most two and therefore
the secant variety $\sigma(\PP^m\times\PP^n)\subset \PP(\CC^{m+1}\otimes\CC^{n+1})$ can be interpreted as the projectivization of the
 locus of rank at most two matrices. When taking higher secants we get 
the well known stratification of bipartite systems by their rank\cite{hey,HLT}
\[\PP^m\times\PP^n\subset \sigma_2(\PP^m\times\PP^n)\subset\dots\subset\sigma_{min(m,n)-1}(\PP^m\times\PP^n)\subset \PP(\CC^{m+1}\otimes\CC^{n+1}).\]
In $\PP^{15}=\PP(\mathcal{H})$ the presence of the three varieties $Seg_i(\PP^3\times\PP^3)$ will induce a first stratification of the ambient space.
Recall the binary shortened notation $|\varphi\rangle=\sum_{i,j,k,l} a_{ijkl}|ijkl\rangle=\sum_{m=0} ^{15} a_m|ijkl\rangle$ where $m=1^i+2^j+4^k+8^l$.
As mentioned in Section \ref{adjoint}, we can represent a four-qubit state $|\varphi\rangle=(a_0,a_1\dots, a_{15})$ by  
a $4\times 4$ matrix \begin{equation}\mathcal{M}_1=\begin{pmatrix}
                                                                                                            a_0 & a_1 & a_2 & a_3\\
a_4 & a_5 & a_6 & a_7\\

a_8 & a_9 & a_{10} & a_{11}\\
a_{12} & a_{13} & a_{14} & a_{15}
                                                                                                           \end{pmatrix}\in\mathcal{H}\end{equation}
%
Similarly we have two alternative embeddings.
\begin{equation}
\mathcal{M}_2=\begin{pmatrix}
                                                                                                            a_0 & a_8 & a_2 & a_{10}\\
a_1 & a_9 & a_3 & a_{11}\\
a_4 & a_{12} & a_{6} & a_{14}\\
a_{5} & a_{13} & a_{7} & a_{15}
                                                                                                           \end{pmatrix}, \mathcal{M}_3=\begin{pmatrix}
                                                                                                            a_0 & a_1 & a_8 & a_9\\
a_2 & a_3 & a_{10} & a_{11}\\
a_4 & a_5 & a_{12} & a_{13}\\
a_{6} & a_{7} & a_{14} & a_{15}
                                                                                                           \end{pmatrix}\end{equation}
%
%
%

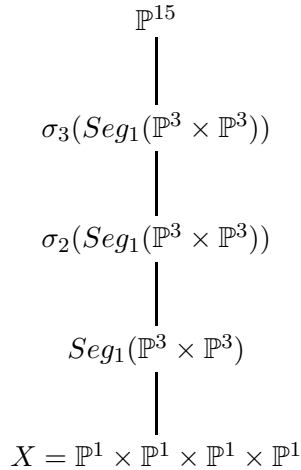
\begin{figure}[!h]
  \[\xymatrix{&\PP^{15}\incl[d]\\
 &\sigma_3(Seg_1(\PP^3\times\PP^3))\incl[d]\\
&\sigma_2(Seg_1(\PP^3\times\PP^3))\incl[d]\\
&Seg_1(\PP^3\times\PP^3)\incl[d]\\ 
& X=\PP^1\times\PP^1\times\PP^1\times\PP^1
}\]
\caption{Inclusion diagram of $X$ within the stratification defined by $\phi_1$}\label{segre}
\end{figure}

If we think in terms of matrix rank, it is clear  that the variety $Seg_i(\PP^3\times\PP^3)$  is defined by the zero locus of the two by two minors of 
the matrix $\mathcal{M}_i$, the secant variety
 $\sigma(Seg_i(\PP^3\times\PP^3))$ is defined by the zero locus of the $3\times 3$ minors of $\mathcal{M}_i$ and 
the third secant variety $\sigma_3(Seg_i(\PP^3\times\PP^3))$ is defined by the vanishing of $\text{det}(\mathcal{M}_i)$ (Figure \ref{segre}). 

Table \ref{byrank} shows how the normal forms, following Verstraete {\em et al.}'s notation, fit in this stratification.
\begin{table}
\begin{tabular}{|c|c|c|c|}
\hline
 Varieties& Forms & rank \\
 \hline
 $\PP^{15}\smallsetminus \sigma_3(\PP^3\times\PP^3)$ & $G_{abcd}$, $L_{abc_2}$, $L_{ab_2}$  &  $4$\\
                                                 
                                                    & $L_{ab_3}$, $L_{a_4}$ & \\
       \hline
$\sigma_3(\PP^3\times\PP^3)$ & $L_{a_20_{3\oplus \overline{1}}}$, $L_{0_{5\oplus\overline{3}}}$, $L_{0_{7\oplus \overline{1}}}$ &$3$\\                                         
                                    
                                                   &$G_{0bcd}, G_{a0cd}, G_{ab0d}, G_{abc0}$               & \\
                                                   & $L_{0bc_2}$, $L_{a0c_2}$,  $L_{ab0_2}$&\\
                                                   & $L_{0b_2}$,  $L_{a0_2}$ & \\
                                                   & $L_{0b_3}$, $L_{a0_3}$ & \\
                                                   & $L_{0_4}$ &\\
                                                    \hline
 $\sigma_2(\PP^3\times\PP^3)$ &  $L_{0_{3\oplus \overline{1}}0_{3\oplus \overline{1}}}$& $2$\\ 
                              & $G_{00cd}, G_{0b0d}, G_{0bc0},G_{a00d},G_{a0c0},G_{ab00} $                                                       & \\
                      & $L_{00c_2}$, $L_{a00_2}$,  $L_{0b0_2}$&\\
                      & $L_{00_2}$&\\
                       & $L_{00_3}$  &\\
                       \hline
      $\PP^3\times\PP^3$  & $G_{a000}, G_{0b00}, G_{00c0}, G_{000d}$                                                       &  $1$\\   
       & $L_{000_2}$ &\\
       \hline
\end{tabular}
\caption{Stratification of the ambient space by rank of $S_1$}\label{byrank}
\end{table}

One can obtain similar tables for the stratification by rank of $\mathcal{M}_2$ and $\mathcal{M}_3$. The corresponding forms are obtained by permuting the qubits by the permutations
$1324$ and $1432$.

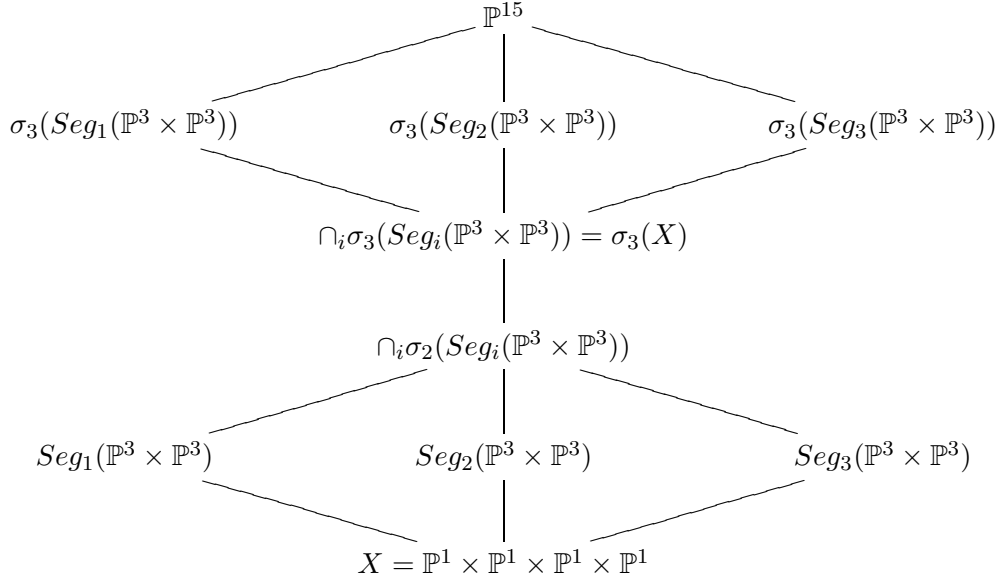
\begin{figure}[!h]
  \[\xymatrix{&\PP^{15}\incl[d]\incl[dr]\incl[dl]& \\
 \sigma_3(Seg_1(\PP^3\times\PP^3))\incl[dr] &\sigma_3(Seg_2(\PP^3\times\PP^3))\incl[d]&\sigma_3(Seg_3(\PP^3\times\PP^3))\incl[dl] \\
&\cap_{i}\sigma_3(Seg_i(\PP^3\times\PP^3))=\sigma_3(X)\incl[d]& \\
&\cap_i \sigma_2(Seg_i(\PP^3\times\PP^3))\incl[d]\incl[dr]\incl[dl]\\ 
Seg_1(\PP^3\times\PP^3)\incl[dr]  & Seg_2(\PP^3\times\PP^3)\incl[d] & Seg_3(\PP^3\times\PP^3)\incl[dl]\\
& X=\PP^1\times\PP^1\times\PP^1\times\PP^1
}\]
\caption{Inclusion diagram of $X$ within the stratification defined by multirank}\label{fmultirank}
\end{figure}

We can do a little bit better by using simultaneously the strafication by rank of the three embeddings of $\PP^3\times\PP^3$. It leads to a stratification by multirank (Figure \ref{fmultirank}, Table \ref{byrank}).

\begin{table}
\begin{tabular}{|c|c|c|}
\hline
 Varieties& Forms &  mutli-rank \\
 \hline
 $\PP^{15}\smallsetminus \cup_{i\in\{1,2,3\}}\sigma_3(Seg_i(\PP^3\times\PP^3))$ & $G_{abcd}$ &  $[4,4,4]$\\
                                                    & $L_{abc_2}$ & \\
                                                   & $L_{ab_3}$ & \\
                                                   \hline
    $\sigma_3(Seg_1(\PP^3\times\PP^3))$     &   $L_{a_2b_2}^{4231}$,    $L_{a_4}^{4231}$       &  $[3,4,4]$\\  
    \dotfill & \dotfill &\dotfill  \\
 $\sigma_3(Seg_2(\PP^3\times\PP^3))$                   & $L_{a_2b_2}$, $L_{a_4}$  &  $[4,3,4]$\\  
 
 \dotfill & \dotfill &\dotfill  \\
 
 $\sigma_3(Seg_2(\PP^3\times\PP^3))$ & $L_{a_2b_2}^{2134}$, $L_{a_4} ^{2134}$  &  $[4,4,3]$ \\

\hline
$\cap_{i\in \{1,2,3\}}\sigma_3(Seg_i(\PP^3\times\PP^3))$ & $L_{a_20_{3\oplus1}}, L_{0_{7\oplus\overline{1}}}$ &$[3,3,3]$\\                                         
         $=\sigma_3(X)$                                           &  &\\

                                                    \hline
 $\cap_{i\in \{1,2,3\}}\sigma_2(Seg_i(\PP^3\times\PP^3))$ &  $L_{0_{3\oplus \overline{1}}0_{3\oplus \overline{1}}}$& $[2,2,2]$\\     
    \hline
\end{tabular}
\caption{Stratification of the ambient space by multi-rank (only general values of the parameters are taken into account).}\label{byrank}
\end{table}

Let us point out that the variety $\cap_{i\in \{1,2,3\}}\sigma_3(Seg_i(\PP^3\times\PP^3))$ is nothing but $\sigma_3(\PP^1\times\PP^1\times \PP^1\times\PP^1)$ the third secant variety of the set 
of separable states $X$ which was described in our previous article\cite{HLT2}. 

The stratification of entanglement classes of four-qubit states by multirank has been already studied in Ref\cite{Cao} but without the geometric interpretation in terms of secants of Segre varieties. Table \ref{byrank} is 
identical to Table 4 of Ref\cite{Cao}.

Finally let us point out that the stratification by rank of the hypersurface $\{L=0\}$ (respectively $\{M=0\}$ and $\{N=0\}$) 
corresponds to a stratification by singular locus of the hypersurface. In the next section we will see that the study of the singular locus of the hypersurface $\Delta=0$ is more challenging. 

\subsection{Singularities of the dual variety $X^*$}
In this section we consider the dual variety of $X$ denoted by $X^*$ and given by the zero locus of the hyperdeterminant $\Delta$ of format $2\times 2\times 2\times 2$. 
This invariant polynomial of four qubits is of fundamental importance in the study of four-qubit states. Its importance was first emphazised by Miyake\cite{My} and recently Gour and Wallach\cite{GW}
used $|\Delta|$ as an entanglement measure of genuine four-qubit entanglement. It is well-known that the nonvanishing of $\Delta$ characterizes semi-simple elements for the $\SLOCC$-action 
on $\mathcal{H}$. The SLOCC stratification of $X^*=\{|\varphi\rangle \in \PP(\mathcal{H}) : \Delta(|\varphi\rangle)=0\}$ by invariant subvarieties has been regarded in Ref\cite{My} in the context of QIT based 
on the earlier work of Weyman and Zelevinsky\cite{WZ}. The purpose of the section is to complete the picture by a finer grained stratification of $X^*$ based on the study of the 
hyperplane sections 
of $X$ and the understanding of this stratification in terms of normal forms and vanishing of invariants.


A hyperplane $H$ belongs to $X^*$ if the corresponding $H$ is tangent to $X$ at some smooth point $x\in X$, i.e. $\tilde{T}_x X\subset H$. In other words the 
hyperplane section $X\cap H$ defines a singular hypersurface of $X$ with a singular point at $x$.
 The singularity, i.e. the hyperplane section with a singular point $x$,  will be denoted by $(X\cap H,x)$.
When $X^*$ is a hypersurface and $H$ a smooth point of $X^*$, it is well-known that the singular hypersurface $X\cap H$ has a unique 
singular point and the Hessian matrix at this unique singular point is  nondegenerate\cite{GKZ}. For a hyperplane section $X\cap H$ we will denote by $f_{X\cap H}$ the polynomial defining $X\cap H$ as a hypersurface of $X$. Therefore $H$ is a smooth point of 
$X^*$ reads as there exists a unique $x\in X$ such that $f_{X\cap H}(x)=0$, $\partial_i f_{X\cap H}(x)=0$ and $Hess(f_{X\cap H},x)$ is of full rank.
Such singular point is called a Morse singularity or a $A_1$ singularity in Arnold's classification of simple singular germs\cite{A}. 
Thus if $H$ is a smooth point of $X^*$, there exists a unique $x\in X$ such that  $(X\cap H,x)\sim A_1$.
When $X^*$ is a hypersurface, 
a singular point of $X^*$ is a hyperplane whose corresponding hyperplane section $X\cap H$ does not have a unique $A_1$ singularity. Therefore there are two possibilities to not satisfy this condition:
\begin{itemize}
 \item Either $X\cap H$ has more than one singular point,
 \item or $X\cap H$ has a unique singular point and $Hess(f_{X\cap H},x)$ is not of maximal rank.
\end{itemize}

This leads to the notion of node and cusp components of the singular locus of $X^*$ as defined in Ref\cite{WZ}.

\begin{def }
 Let $X\subset \PP(V)$ a nondegenerate projective variety and $X^*$ its dual variety which is assumed to be a hypersurface. 
 The singular locus $\text{Sing}(X^*)$ of $X^*$ is given by 
 \begin{equation}
  \text{Sing}(X^*)=X^*_{node}\cup X^*_{cusp}
 \end{equation}

 where the node component, $X^*_{node}$ is defined by 
 \begin{equation}
  X^* _{node}=\overline{\{H\in X^*:  \exists(x,y)\in X\times X, x\neq y, \tilde{T}_x X\subset H, \tilde{T}_y X\subset H\}}
 \end{equation}
 and the cusp component, $X^*_{cusp}$ is
 \begin{equation}
  X^*_{cusp}=\overline{\{H\in X^*:  \exists x\in X, \tilde{T}_xX\subset H, (X\cap H,x)\not\sim A_1\}}.
 \end{equation}
\end{def }

As pointed out in Ref\cite{WZ}, in the case where $X$ is a Segre product, the node component may be further decomposed. We need to introduce the notion of $J$-node component.
We give the definition in the case of a product of projective spaces but it can easily be extended to a Segre product of algebraic varieties.

\begin{def }
 Let $X=\PP^{k_1}\times \dots\times \PP^{k_r}\subset \PP^{(k_1+1)\dots(k_r+1)-1}$ be the Segre product of $r$ projective spaces. Let $J=\{j_1,\dots,j_s\}\subset \{1,\dots,r\}$.
 We say that $(x,y)\in X\times X$ is a $J$-pair of points when $x=x_1\otimes x_2 \otimes \dots \otimes x_{j_1}\otimes \dots\otimes x_{j_r}\otimes \dots\otimes x_{r}$ and 
 $y=y_1\otimes y_2 \otimes \dots \otimes x_{j_1}\otimes \dots\otimes x_{j_r}\otimes \dots\otimes y_{r}$. Then the $X^*_{node}(J)$ singular locus is defined  by 
 \begin{equation}
  X^*_{node}(J)=\overline{\{H\in X^*:\exists (x,y) \text{ a }J\text{-pair of points of } X\times X, \tilde{T}_x X\subset H, \tilde{T}_y X\subset H\}}.
 \end{equation}

\end{def }
In Ref \cite{WZ} it is proven that for $X=\PP^1\times\PP^1\times \PP^1\times \PP^1$, the irreducible components of $\text{Sing}(X^*)$ are
\begin{equation}
 \text{Sing}((\PP^1\times\PP^1\times \PP^1\times \PP^1)^*=X^*_{cusp}\cup X^*_{node}(\emptyset)\cup\bigcup_{1\leq i\leq j\leq 4} X^* _{node}(\{i,j\})
\end{equation}

Cusp and node components have interpretation in terms of duals of auxiliary varieties of $X$. In the case of four-qubit systems one has the following proposition.
\begin{prop}\label{cusp}
 Let $X=\PP^1\times\PP^1\times\PP^1\times \PP^1$, then
 \begin{enumerate}
  \item $X^* _{node}=\sigma(\PP^1\times\PP^1\times\PP^1\times \PP^1)^*$
  \item $X^* _{node}(\{i,j\})= \sigma_{\{i,j\}}(\PP^1\times\PP^1\times\PP^1\times\PP^1)^*$, 
  \item $X^* _{cusp}=\tau(\PP^1\times\PP^1\times\PP^1\times\PP^1)^*$
 \end{enumerate}
where $\sigma_{\{J\}}(\PP^1\times\PP^1\times\PP^1\times\PP^1)$ is the secant variety of $J$-pairs of points, i.e. $\sigma_{\{J\}}(\PP^1\times\PP^1\times\PP^1\times\PP^1)=\ds\overline{\bigcup_{(x,y) J\text{-pair}} \PP_{xy}}$ .
\end{prop}
\proof The points 1 and 2 are already particular cases of  Proposition 4.1 of our previous paper\cite{HLT} where we establish that the node component is always the dual variety 
of some $J$-secant variety of the original Segre product. The $J$-secant variety is the variety of secant lines where the lines are defined by $J$-pairs 
of points. The proof
of this general statement follows from the application of the Terracini's lemma (see Ref\cite{HLT}).

Point 3 is more subtle. First it should be noticed that $\tau(X)^*\subset X^*_{cusp}$. This can be understood from the fact that a hyperplane $H$ tangent to $\tau(X)$ is also 
tangent to $X$. Moreover the fact that $H$ is tangent to $\tau(X)$ at $v$ implies that $H$ is tangent to $X$ along the direction $v$. Therefore the matrix $Hess(f_{X\cap H},x)$ is degenerate
in the direction $v$. This implies 
that $\tau(X)^*\subset X^* _{cusp}$.
The equality will follow if we prove equality of dimension.
Thus one has to calculate  the dimension of $\tau(\PP^1\times\PP^1\times\PP^1\times\PP^1)^*$. The dimension of a dual variety can be calculated by 
Katz's formula (see Ref\cite{GKZ}) which states that for a projective variety $Y\subset \PP(V)=\PP^N$, the dimension of $Y^*$ is obtained from
\begin{equation}
 \text{dim}(Y^*)=N-\text{min}_{H\in X^*}(\text{corank}(Hess(f_{Y\cap H},x))-1
\end{equation}
where $x$ is the point where $f_{Y\cap H}$ is singular. In particular this formula says that in general, the dual variety is a hypersurface, because we expect 
$\text{min}_{H\in X^*}(\text{corank}(Hess(f_{Y\cap H},x))$ to be zero (when it is not, it means that all tangent hyperplanes are tangent to $Y$ not to a point but to a subspsace of positive dimension).
This formula can be used to compute the dimension of $\tau(X)^*$. For this purpose, one needs to compute the general form of $Hess(f_{\tau(X)\cap H},x)$. 
Assuming $\text{dim}\sigma(X)=2n+1$ (and thus $\text{dim}\,(\tau(X))=2n$ see Ref\cite{Z2}), it can be shown using moving frames techniques that 
\begin{equation}
 Hess(f_{\tau(X)\cap H},v)=\begin{pmatrix}
                            A & Hess(f_{X\cap H},x) \\
                            Hess(f_{X\cap H},x) & 0\\
                           \end{pmatrix}
\end{equation}
where $v$ is a general point of $\tau(X)$ (i.e. a general element of $\tilde{T}_x X$) and $A$ is a $n\times n$ full rank block  built from the cubic invariants of the Taylor expansion of 
$f_{X\cap H}$ at $x$. Because $H$ belongs to $\tau(X)^*$ it is tangent to $X$ along the direction $v$ and necessarly $\text{rank}(Hess(f_{X\cap H},x)\leq n-1$ confirming the fact that 
$\tau(X)^*$ is at most of codimension $2$ in the ambient space. To prove that $\tau(X)^*$ is, in our case, of codimension $2$, one needs to show that 
we can find a matrix $Hess(f_{X\cap H},x)$ which is of rank $n-1$ for a generic tangent vector $v$. In our situation $X=\PP^1\times \PP^1\times\PP^1\times\PP^1$ and 
let us assume $x=|0000\rangle$. Let  $v=|1000\rangle+|0100\rangle+|0010\rangle+|0001\rangle$ be a generic tangent vector to $X$ at $x$, and  consider a  general 
curve $t\to\gamma(t)$, such that $\gamma(0)=x$ and $\gamma'(0)=v$. 
Then we have
\begin{equation}
\begin{split}
 \gamma(t)=|0000\rangle+t(|1000\rangle+|0100\rangle+|0010\rangle+|0001\rangle)\\
 +\dfrac{t^2}{2!}(|1100\rangle+|1010\rangle+|1001\rangle+|0110\rangle+|0101\rangle+|0011\rangle)+O(t^3).
 \end{split}
\end{equation}
For the hyperplane $a\langle1100|+b\langle1010|+c\langle 1001|+d\langle0110|+e\langle0101|+f\langle 0011|$, we obtain 
\begin{equation}
 Hess(f_{X\cap H},x)=\begin{pmatrix}
                      0 & a & b & c\\
                      a & 0 & d & e\\
                      b & d & 0 & f\\
                      c & e & f & 0\\
                     \end{pmatrix}.
\end{equation}

One can check that for $a=f$, $b=e$, $d=c=-e-f$, then $v$ is a singular direction of $Hess(f_{X\cap H},x))$ and  $\text{rank}(Hess(f_{X\cap H},x))=3$. Therefore $\text{dim}(\tau(X)^*)=13$ and the equality follows by irreducibility of $X^* _{cusp}$. $\Box$

Proposition \ref{cusp} gives a description of the singular locus of the hyperdeterminants in terms of the dual varieties of (the orbit closure) of GHZ and W states for the four-qubit systems. 
Combining the description of $\text{Sing}(X^*)$ established in Ref\cite{WZ} and our interpretation of the singular components in terms of  tangential and secant varieties we get,

\begin{theorem}
 Let $\mathcal{H}=(\CC^2)^{\otimes n}$ the Hilbert space of a $n$-qubit system with $n\geq 3$. Let $X=\PP^1\times\dots\times\PP^1$ be the set of separable states and $X^*$ 
 its dual variety given by the vanishing of the $2\times \dots\times 2$ hyperdeterminant.
 Then we have
 \begin{enumerate}
  \item $\text{Sing}(X^*)=\sigma_{\{1\}}(X)^*\cup \sigma_{\{2\}}(X)^* \cup \sigma_{\{2\}}(X)^*$ for $n=3$,
  \item $\text{Sing}(X^*)=\tau(X)^*\cup \sigma(X)^*\cup_{1\leq i<j\leq 4}\sigma_{\{i,j\}}(X)^*$ for $n=4$,
  \item $\text{Sing}(X^*)=\tau(X)^*\cup \sigma(X)^*$ for $n\geq 5$.
 \end{enumerate}
\end{theorem}

\proof Point 1 is already proved in our paper Ref\cite{HLT} where the $3$-qubits case is studied in details. Point 2 is Proposition \ref{cusp}. Now to prove
point 3 we use the result of Weyman and Zelevinsky which states that in 
this case $\text{Sing}X^*=X^*_{cusp}\cup X^*_{node}(\emptyset)$. Weyman and Zelevinsky also proved that both components are irreducible and of codimension $1$ in $X^*$. 
The identification $\sigma(X)^*=X^*_{node}(\emptyset)$ follows from  Terracini's lemma. The inclusion $\tau(X)^*\subset X^*_{cusp}$ is clear from the description of $\tau(X)$ but to prove the equality one needs to calculate the 
dimension of $\tau(X)^*$. The same argument as Proposition \ref{cusp} shows that it is equivalent to find a $n\times n$ symmetric matrix $Q$  of rank $n-1$, with $0$ on the diagonal, such that $\ker\ Q= <(1,\dots,1)>$.
Such a matrix can be constructed as follow:
\begin{itemize}
 \item If $n=2k$ we consider the matrix $Q$ defined by $q_{ii}=0, q_{i,n+1-i}=-2(k-1)$ and $q_{i,j}=1$ if $(i,j)\notin \{(i,i),(i,n+1-i)\}$.
 \item If $n=2k+1$ we consider the matrix $Q$ defined by $q_{ii}=0, q_{i,n+1-i}=-2k+1$ for $1\leq i\leq k$ and $k+2\leq i\leq n$, $q_{k+1,k}=q_{k+1,k+2}=-k$ and $q_{i,j}=1$ for 
 $(i,j)\notin \{(i,i),(i,n+1-i),(k+1,k),(k+1,k+2)\}$.
\end{itemize}

 In both cases the matrix $Q$ is symmetric with rank $n-1$ and $\ker\ Q_n=<(1,\dots,1)>$. Therefore one can construct a hyperplane section of $\tau(X)$ such that the corresponding Hessian is of rank $2n-1$, which implies that 
 $\tau(X)^*$ is of codimension $1$ in $X^*$. $\Box$.
\begin{rem}\rm
 It is interesting to point out that the two main stratas of $\text{Sing}(X^*)$ are dual varieties of the orbit closure of GHZ and W-states. 
\end{rem}
                                  
The cusp and node components can be further decomposed by their multiplicity. Let us consider 
\begin{equation}
 X^* _n=\overline{\{H\in X^*: \text{mult}_H X^*=n\}}, \text{ i.e } H \text{ is a root of multiplicity }n \text{ of }\Delta
\end{equation}
Then we have a filtration of $X^*$ by mutliplicities: $X^*\supset X^* _1\supset X^* _2\supset\dots\supset \dots$.

We can define
\begin{equation}
 X^* _{node,k}(J)=X^* _{node}(J)\cap X^* _k
\end{equation}
and similarly 
\begin{equation}
 X^* _{cusp,k}=X^* _{cusp}\cap X^* _k.
\end{equation}

A result of Dimca\cite{Dimca}, generalized by Parusinski\cite{Paru}, shows that the multiplicity of a hyperplane $H\in X^*$ is equal to the Milnor number of the hyperplane section $X\cap H$. 
The Milnor number of an isolated singularity $(f_{X\cap H},x)$ is a topological invariant  defined by 
\begin{equation}
 \mu=\dim_\CC \mathcal{O}_k /(\nabla f)
\end{equation}
where $\mathcal{O}_k$ is ring of all germs $g:(\CC^k,x)\to (\CC,0)$ and $(\nabla f)=(\dfrac{\partial f}{\partial x_1}(x),\dots,\dfrac{\partial f}{\partial x_k}(x))$ is the gradient ideal.
If $X\cap H$ has only isolated singularities, the Milnor number of the hyperplane section $X\cap H$, denoted by $\mu(X\cap H)$, will be the sum of the Milnor number of each singularity. 
Under this assuption, the result of Dimca says
\begin{equation}
 mult_{H}(X^*)=\mu(X\cap H).
\end{equation}
Which, geometrically, leads to the following observation of F. Zak\cite{Z3}
\begin{equation}
 \sigma_n(X)^*\subset X_n^*.
\end{equation}

In Ref\cite{HLP} we have calculated the isolated singular types of the hyperplane sections of the set of separable states 
of four qubits, $X=\PP^1\times\PP^1\times\PP^1\times\PP^1$, for all possible linear forms (hyperplane) obtained by Verstraete {\em et al.}. The construction we employed was the following, let $|\varphi\rangle$ be a state given by Verstraete {\em et al.} classification
and let us consider the linear form $\langle \varphi|$. Then the hyperplane section $H_{\langle \varphi|}\cap X$ defines a hypersurface of $X=\PP^1\times\PP^1\times\PP^1\times\PP^1$ which may be smooth or 
have singularities. When the singularities are isolated, we have tools coming from the classification of simple singularities \cite{A} to discriminate the corresponding hyperplane sections.
The type of the  hypersurface $X\cap H\subset X$ is SLOCC invariant (because $X$ is a homogeneous SLOCC orbit) and so is the singularity type attached to $X\cap H$.

\subsection{Geometric stratification of the Hilbert space of four-qubit states}
The singular type of a hyperplane section $X\cap H$ can be used to discuss the different stratas of the hyperdeterminant $\Delta$ as we now explain. 
For instance it is well known that, for generic $a,b,c,d$,
the states $G_{abcd}$ are such that $\Delta(G_{abcd})\neq 0$, which means that $H_{G_{abcd}}\cap X$ is a smooth hypersurface. By calculating\cite{HLP} for each
Verstraete {\em et al.} normal forms, the different types of the isolated singularities, we
 identify
which forms belong to which components of the singular locus of $\Delta$. If the form gives a hypersurface with only one isolated 
singularity of type $A_1$ then the tested form is a smooth point 
of the hyperdeterminant. If the form gives several $A_1$ singularities it is a point of the node locus. If the singularity is not of type $A_1$ it is a point of the cusp locus.
Moreover the Milnor number of the singularity gives information on the multiplicity of the component.

Once the normal forms are interpreted as components of some specific singular locus then we can test the normal forms on the invariant obtainted in Figure \ref{VDiag} 
to 
identify geometrically the varieties obtained from the analysis of the three quartics. 
For instance if we consider the Verstraete form $L_{ab_2}$, the corresponding hyperplane section $H_{\langle L_{ab_2}|}$ has a unique $A_2$ singularity.
It is the only form to have this property (for a generic choice of parameters) and thus the states $|L_{ab_2}\rangle$ form an open subset of $\tau(X)^*$. 
We easly check that $\ket{L_{ab_2}}$ corresponds to the vanishing of $I_2$ and $I_3$, i.e. following the notations of Figure \ref{VDiag} we have $\mathcal{V}_{I_2,I_3}=\tau(X)^*$.

The singular type of the hyperplane section $H_{\bra{\varphi}}\cap X$ being invariant under permutation of the qubits, the forms are given up to a permutation, i.e. $\ket{L_{ab_2}^\sigma}\in \tau(X)^*$.
Analysing similarly all forms of the $9$ families we obtain:
\begin{theorem}\label{theo}
 The Hilbert space of four-qubit states can be stratified under SLOCC according to the hyperdeterminant as shown in Figure \ref{figtheo2}
 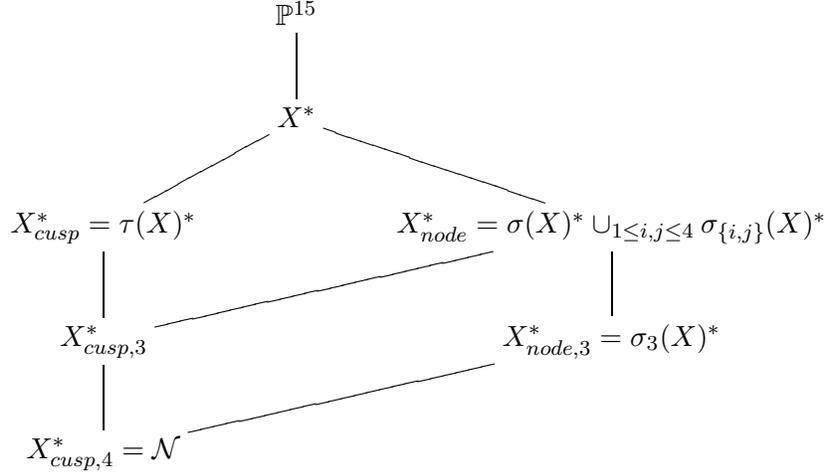
\begin{figure}[!h]
 \begin{equation*}
  \xymatrix{
  & \PP^{15}& \\
  &  X^*\incl[u]  & \\
  X^* _{cusp}=\tau(X)^*\incl[ur] & &X^*_{node}=\sigma(X)^*\cup_{1\leq i,j\leq 4} \sigma_{\{i,j\}}(X)^* \incl[ul]\\
  X^* _{cusp,3}\incl[u] \incl[urr]& & X^*_{node,3}=\sigma_3(X)^*\incl[u] \\
  X^*_{cusp,4}=\mathcal{N}\incl[u]\incl[urr]
  }
 \end{equation*}
 \caption{Stratification of the dual variety of the separable states}\label{figtheo2}
 \end{figure}
with the varieties of the stratification being described in terms of forms and invariants by Tables \ref{tabledd}, \ref{tabledd2}, \ref{tabledd3}.

The stratification can be completed by the stratification of the varieties $Seg_i(\PP^3\times\PP^3)^*$ (Figure \ref{ambient}).
\begin{figure}[!h]
 \begin{equation*}
  \xymatrix{
  & \PP^{15}& \\
  &  X^*\incl[u]  & Seg_1(\PP^3\times\PP^3)^*\incl[ul] & Seg_2(\PP^3\times\PP^3)^*\incl[ull] & Seg_3(\PP^1\times\PP^3)^*\incl[ulll]\\
  X^* _{cusp}=\tau(X)^*\incl[ur] & &X^*_{node}=\sigma(X)^*\cup_{1\leq i,j\leq 4} \sigma_{i,j}(X)^* \incl[ul] \incl[u]\incl[ur]\incl[urr]& \incl[u]\incl[ul]\incl[ur]\sigma_3(X)\\
  X^* _{cusp,3}\incl[u] \incl[urr]& & X^*_{node,3}=\sigma_3(X)^*\incl[u]\incl[ur] \\
  X^*_{cusp,4}=\mathcal{N}\incl[u]\incl[urr]
  }
 \end{equation*}
 \caption{Stratification of the ambient space by SLOCC varieties}\label{ambient}
 \end{figure}
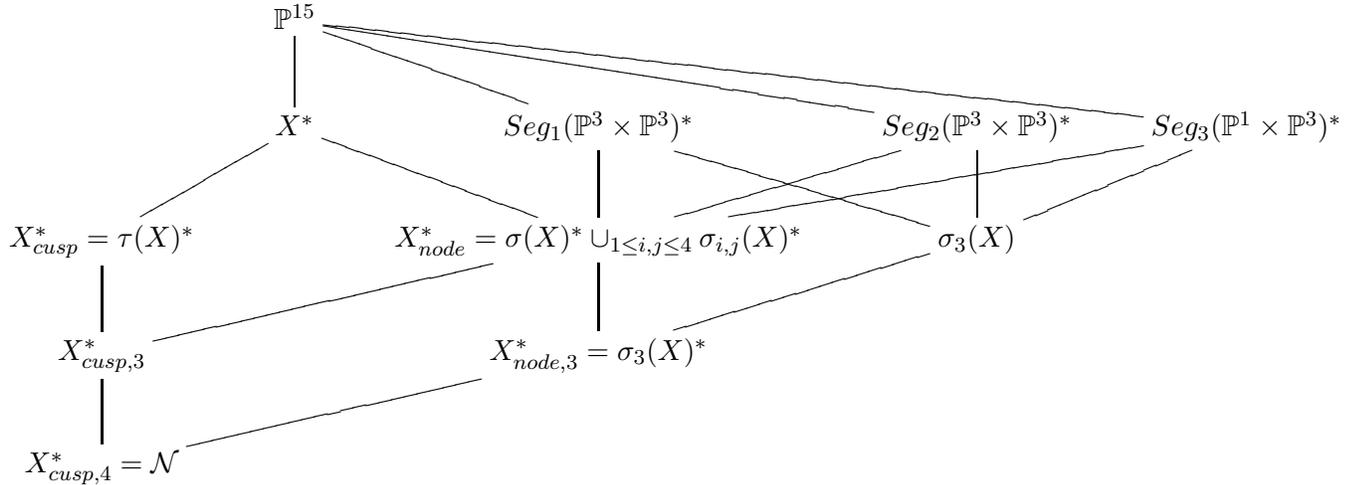
  \begin{table}[!h]
  \begin{tabular}{|c|c|c|c|}
  \hline
   Varieties & Forms & Invariants & Singularities\\
   \hline
    $\PP^{15}\smallsetminus X^*$ & $G_{abcd}$ & $\Delta\neq 0$& smooth hyperplane section\\
     
                                                    \hline
      $X^*$ & $L_{abc_2}$ &$\Delta=0$& hyperplane sections \\
       &               && with a unique $A_1$\\
\hline    
\end{tabular}
\caption{Stratification by $\Delta=0$ (smooth points)}\label{tabledd}
 \end{table}
\begin{table}[!h]
  \begin{tabular}{|c|c|c|c|}
  \hline
   Varieties & Forms & Invariants & Singularities\\
   \hline   
$X^*_{cusp}=\tau(X)^*$ & $L_{ab_3}$ & $I_2=I_3=0$ & hyperplane sections\\
                       &   &                 &  with a unique $A_2$\\
                       \hline
$X^*_{cusp,3}$         & $L_{a_4}$ & $L=P=S_1=0$                & hyperplane sections\\
                        &         &   or $M=D_{xy}=S_2=0$ or $N=D_{xy}=S_3$               & with a unique $A_3$\\
                        \hline
$X^*_{cusp,4}$         & $L_{0_{7\oplus\overline{1}}}$&$B=L=M=D=0$ & hyperplane sections\\
                       &                              & & with a unique $D_4$\\
                       \hline
                       
  \end{tabular}
\caption{Stratification by $\Delta=0$ (cusp components)}\label{tabledd2}
\end{table}
\begin{table}[!h]
  \begin{tabular}{|c|c|c|c|}
  \hline
   Varieties & Forms & Invariants & Singularities\\
   \hline   
$X^*_{node}(\emptyset)=\sigma(X)^*$ & $G_{abcc}$ & see Remark \ref{Stru} & hyperplane sections\\
                       &   &      &  with two  $A_1$ singularities\\
                       \hline
$\cup_{1\leq i<j\leq 4} \sigma_{\{i,j\}}(X)^*$         & $L_{a_2b_2}$ &   $M=D_{xy}=0$             & hyperplane sections\\
                         &         &   or  $N=D_{xy}=0$              & with two $A_1$\\
                  &    &  or $L=P=0$                                  &\\
                        \hline
$X^*_{node,3}$         & $L_{a_20_{3\oplus\overline{1}}}, L_{aac_2}$ & $L=M=D_{xy}=0$                & hyperplane sections\\
                       &                              & & with three $A_1$\\

                              \hline
  \end{tabular}
\caption{Stratification by $\Delta=0$ (node components)}\label{tabledd3}
 \end{table}
 
 \end{theorem}

 \begin{rem}\rm
  The study of the singularities of $\Delta$ enables us to caracterize the Verstraete normal forms as general points of specific stratas. It was well known that $\ket{G_{abcd}}$, for a generic choice of parameters, 
  will be a state on which $\Delta$ does not vanish. Now, one can see that the other forms, for a generic choice of parameters and up to permutation, are general points of specific stratas of the singular locus.
  The two nilpotent states $L_{0_{5\oplus \overline{3}}}$ and $L_{0_{3\oplus\overline{1}}0_{3\oplus \overline{1}}}$ which do not appear in Theorem \ref{theo} belong to the nullcone $\mathcal{N}$.
 \end{rem}

\begin{rem}\rm
 The varieties, as denoted in the Theorem, are not all irreducible. For instance $X^* _{cusp,3}$ is not irreducible and neither is $X_{cusp,4} ^*$ which corresponds to the nullcone $\mathcal{N}$.
\end{rem}

\begin{rem}\label{Stru}\rm
 All varieties of Figure \ref{ambient} correspond to varieties detected in Figure \ref{VDiag}  by the invariants of the three quartics, except $\sigma(X)^*$ which 
 does not appear in Figure \ref{VDiag}  but which can be detected by computing the hyperplane sections of $\ket{G_{abcc}}$. The defining equations 
 of $\sigma(X)^*$ have been computed by Lin and Strumfels\cite{LS}. This component corresponds to the projective closure of the image of the principal minor map for $4\times 4$ matrices.  
\end{rem}

\begin{rem}\rm
 It is interesting to point out that most of the SLOCC varieties exhibited in Figure \ref{ambient} have a quantum information theory interpretation in terms of the duals of 
 well-known quantum states orbit closures. In Figure \ref{dualorbit}, which is a translation of part of Figure \ref{ambient}, we denote by $\overline{|\varphi\rangle}$
 the SLOCC orbit closure of the state $|\varphi\rangle$. We also denote by $Sep$ any separable state and thus  $\overline{|Sep\rangle} =X$. We also denote by $|GHZ_3\rangle$ the family of states which can be written as a sum of three separable states (tensor of rank $3$).
 \begin{figure}[!h]
 \begin{equation*}
  \xymatrix{
  & \PP^{15}& \\
  &  (\overline{|Sep\rangle})^*\incl[u]  & \overline{(|EPR\rangle}\otimes \overline{|EPR\rangle})^*\incl[ul]& \text{ (and its permutations)}\\
  (\overline{|W\rangle})^*\incl[ur] & & (\overline{|GHZ\rangle})^*\cup_{1\leq i,j\leq 4} (\overline{|ij\rangle\otimes |EPR\rangle})^* \incl[ul] \incl[u]&\overline{|GHZ_3\rangle} \incl[ul] \\
  }
 \end{equation*}
 \caption{Stratification of the ambient space by duals of state orbit closures}\label{dualorbit}
 \end{figure}
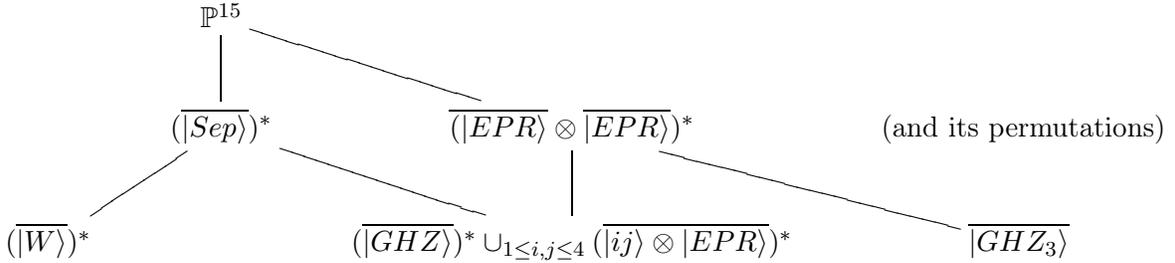
\end{rem}


\section{Verstraete type of a form}\label{algorithme}
Let $\mathbb A$ be a set of variables and $|\varphi(\mathbb A)\rangle$ be a family of parametrized forms where each variable of
 $\mathbb A$ ranges over $\mathbb C$.
A specialization of $\mathbb A$ is a system $S$ of algebraic equations in the variables of $\mathbb A$. We denote by $|\varphi(\mathbb A)\rangle|_{S}$ the subfamilies
of $|\varphi(\mathbb A)\rangle$  such that the values of $\mathbb A$ satisfy $S$. For instance, 
 the four-parameter family $G_{abcd}$ together with $S=\{a^{2}=d^{2},b^2=c^{2}\}$ is union of the two dimensional spaces
generated by one of the four basis states $\{|0000\rangle+|1111\rangle, |0101\rangle+|1010\rangle\}$, 
$\{|0000\rangle+|1111\rangle, |0110\rangle+|1001\rangle\}$, $\{|1100\rangle+|0011\rangle, |0101\rangle+|1010\rangle\}$, or
$\{|1100\rangle+|0011\rangle, |0110\rangle+|1001\rangle\}$.
 
A form $\varphi$ has  a Verstraete type $[F,S]$, where $F$ is one of the nine Verstraete generic forms and $S$ a specialization of the 
parameters, if there exists a permutation $\sigma\in\S_{4}$ of the qubits such that $\varphi$ is SLOCC-equivalent to an element of
 $(F|_{S})^{\sigma}$. 

In this section, we describe an algorithm allowing  to compute the Verstraete type for any given form.
First note that if the form is nilpotent it is easy to find a Verstraete equivalent form from our previous paper \cite{HLT2} and Table \ref{TVN}. 
Now suppose that the form $|\varphi\rangle$ is not nilpotent. We use the 170 covariants computed in our previous paper \cite{HLT2} in order to discriminate between the
Verstraete forms. In particular, we define:
\[
\mathcal L=L_{6000}+L_{0600}+L_{0060}+L_{0006}
\]
\[
\mathcal K_{5}=K_{5111}+K_{1511}+K_{1151}+K_{1115},\ \mathcal K_{3}=K_{3311}+K_{3131}+K_{3113}+K_{1331}+K_{1313}+K_{1133},
\]
\[\overline{\mathcal G}=G_{3111}^{1}G_{1311}^{1}G_{1131}^{1}G_{1113}^{1},\ 
\mathcal G=G_{3111}^{2}+G_{1311}^{2}+G_{1131}^{2}+G_{1113}^{2},\]
\[
\mathcal D=D_{4000}+D_{0400}+D_{0040}+D_{0004},\]
\[\mathcal H=H_{2220}+H_{2202}+H_{2022}+H_{0222}\]
and
$\mathcal C=C_{1111}^{2}$.
We proceed as follows: first we classify the forms with respect to the roots of the three quartics $Q_{1}$, $Q_{2}$ and $Q_{3}$ 
according to the discussion in Appendix \ref{DiscRoots}.
For each of the cases considered in Appendix \ref{DiscRoots}, we determine which Verstraete forms can occur and, when there are several 
possibilities we use one of the covariants previously defined to discriminate between them. Let $V$ be a vector, 
we denote by $ev(V)$ the vector such that $ev(V)[i]=0$ if $V[i]=0$ and $ev(V)[i]=1$ if $V[i]\neq 0$.
\begin{enumerate}
\item If the quartics have only nonzero roots
\begin{enumerate}
\item If all the roots are simple then this is the generic case and the Verstaete type is $[G_{abcd};\emptyset]$.
\item If each quartic has double root and two simple roots (equivalently $T_{1}=T_{2}=T_{3}=0$ and $I_{2},I_{3}\neq 0$ then
two cases can occur. Either the Verstraete type is $[G_{abcd};{c=d}]$ or it is $[L_{abcc};\emptyset]$. We determine the forms 
remarking that $\mathcal L(L_{abc_{2}})\neq 0$ and $\mathcal L(G_{abcc})=0$.
\item If each quartic has a single simple root and a triple root (equivalently $I_{2}=I_{3}=0$) then 
 three cases can occur: $[G_{abcd};b=c=d]$,
$[L_{abc_{2}};b=c]$ and $[L_{ab_{3}};\emptyset]$. In order to determine the type, we evaluate the vector $V=[\mathcal K_{5},\mathcal L]$ on each forms. 
We can decide the type of the form according to the values
\[ev(V(G_{abbb}))=[0,0],\ ev(V(L_{abb_{2}}))=[1,0],\mbox{ and }ev(V(L_{ab_{3}}))=[1,1]\]
\end{enumerate}
\item If only one of the quartics $Q_{i}$ has a zero root then
\begin{enumerate}
\item If $Q_{i}$ has only simple roots then the only possibility is $[G_{abcd};d=0]$
\item If $Q_{i}$ has a double zero root and two simple roots then we have three possibilities $[G_{abcd};c=d=0]$, $[L_{abc_2};c=0]$ 
or $[L_{a_{2}b_{2}};\emptyset]$.{}
We evaluate the form on the vector $V=[\mathcal K_{3},\mathcal L]$ and compare with
\[ev\left(V(G_{ab00})\right)=[0,0],\ ev\left(V(L_{ab0_{2}})\right)=[1,0],\mbox{ and } ev\left(V(L_{a_{2}b_{2}})\right)=[1,1]\]
\item If $Q_{i}$ has triple zero root and a simple root then we have the five possibilities
 $[G_{abcd};c=b=d=0]$, $[L_{abc_{2}};b=c=0]$, $[L_{ab_{3}};a=0]$, $[L_{a_{2}b_{2}};a=b]$ and $[L_{a_{4}};\emptyset]$. 
 We evaluate the form on the vector $V=[\mathcal C,\mathcal D,\mathcal K_{5},\mathcal L]$ and compare to the identities
 \[ev\left(V(G_{a000})\right)=[0,0,0,0],\ ev\left(V(L_{a00_{2}})\right)=[1,0,0,0],\]
 \[ ev\left(V(L_{0b_{3}})\right)=[1,1,1,0],\ 
 ev\left(V(L_{a_{2}a_{2}})\right)=[1,1,0,0]\]
  \ and  $ev\left(V(L_{a_{4}})\right)=[1,1,1,1]$.
 \item If $Q_{i}$ has a double nonzero root and two simple roots then there are two possibilitites 
 $[G_{abcd};b=c,d=0]$ and $[L_{abc_{2}};b=0]$ which can be identified by remarking that $\mathcal L(G_{abb0})=0$ and 
 $\mathcal L(L_{a0c_{2}})\neq 0$.
 \item If $Q_{i}$ has a triple nonzero root then one has to examine 3 possibilities: $[G_{abcd};a=b=c,d=0]$, $[L_{abcc};b=c,a=0]$ and 
 $L_{abbb};b=0]$.
 It suffices to consider the vector $V=[\mathcal D,\mathcal L]$ and remark that
 \[ev(V(G_{aaa0}))=[0,0],\ ev(V(L_{0cc_{2}}))=[1,0],\mbox{ and }ev(V(L_{a0_{3}}))=[1,1].
 \] 
\end{enumerate}
\item If each quartic has at least a zero root then
\begin{enumerate}
\item If all the roots are simple then the type is $[G_{abcd};d=0]$.
\item If all the zero roots are simple and there is a nonzero double root then we have 2 possibilities $[G_{abcd};b=a,c=-2a,d=0]$,
$[L_{ab_{2}};a=0,c=\frac b2]$. We can discriminate between these two cases by remarking $\mathcal L(G_{aa(-2a)0})=0$ and 
$\mathcal L(L_{0(2b)b_{2}})\neq 0$.
\item If all the zero roots are double then we have to consider 5 cases: $[G_{abcd};a=b=0,c=d]$, $[L_{abc_{2}};a=b,c=0]$,
 $[L_{abc_{2}};a=b=0]$, $[L_{a_2b_{2}};a=0]$, and $[L_{a_{2}0_{3\oplus\overline 1}};\emptyset]$. 
 We consider the vector $V=[\overline{\mathcal G},\mathcal G,\mathcal H,\mathcal L]$. The evaluation of this vector on the different cases gives
 \[ev(V(G_{00aa}))=[0,0,0,0],\, ev(V(L_{aa0_{2}}))=[0,1,1,0],\] 
 \[ev(V(L_{00c_{2}}))=[0,0,1,0], ev(L_{0_2a_{2}})=[1,1,1,0],\mbox{ and }
 ev(L_{a_{2}0_{3\oplus\overline 1}})=[1,1,1,1].
 \]
\end{enumerate}
\end{enumerate}
\begin{ex}\rm
	\rm
	Consider the form $|\varphi\rangle=2|0100\rangle+|1101\rangle+4|1111\rangle+3|0010\rangle$.  All the quartics are equal
	 $Q_{1}(|\varphi\rangle)=Q_{2}(|\varphi\rangle)=Q_{3}(|\varphi\rangle)=x^{2}(x+3y)^{2}$. Hence, we are in the case (3.c) of the algorithm. 
	 We compute $ev([\overline{\mathcal G}(|\varphi\rangle),\mathcal G(|\varphi\rangle),\mathcal H(|\varphi\rangle),\mathcal L(|\varphi\rangle)])=[0,1,1,0]$.{}
	 So the type of $|\varphi\rangle$ is $L_{aa0_{2}}$.
\end{ex}

\begin{ex}\rm
	\rm
	Consider the form $|GHZ\rangle=|0000\rangle+|1111\rangle$.  All the quartics are equal
	 $Q_{1}(|\varphi\rangle)=Q_{2}(|\varphi\rangle)=Q_{3}(|\varphi\rangle)=x^{2}(x-y)^{2}$. Hence, we are also in the case (3.c) of the algorithm. 
	 We compute $ev([\overline{\mathcal G}(|GHZ\rangle),\mathcal G(|GHZ\rangle),\mathcal H(|GHZ\rangle),\mathcal L(|GHZ\rangle)])=[0,0,0,0]$.{}
	 So the type of $|\varphi\rangle$ is $G_{00aa}$. Indeed, from the definition $|GHZ\rangle=G_{1001}$.\\
	 In the same way, the product of two EPR states $|\varphi_{1}\rangle=|0000\rangle+|0011\rangle+|1100\rangle+|1111\rangle$ gives $Q_{1}=Q_{3}=(x-y)^{4}$, 
	 and $Q_{2}=x^{3}(x-4y)$. Hence, we are in the case (2.c) and 
	 since $ev([\mathcal C,\mathcal D,\mathcal K_{5,\mathcal L}])=[0,0,0,0]$, this implies that $\varphi_{1}$ has the type $G_{a000}$, as expected. Indeed, it is exactly $2G_{1000}$.
\end{ex}

\begin{ex}\rm
	In Table \ref{TabLM}, we summarize the type of the forms used to describe the inclusion diagram of the third secant variety \cite{HLT2}.
	\end{ex}
	\begin{table}
		\[
		\begin{array}{|c|c|}
			\hline |\varphi\rangle&\mbox{ Type}\\\hline
			65257&G_{abc0}\\
			6014&L_{0(2a)b_{2}}\\
			65261,
			65513,
			65273,65259&L_{a_{2}0_{3\oplus\overline 1}}\\
			59777&G_{abc0}\\
			59510&G_{aa(-2a)0}\\
			65267,
			65509,
			65507,
			65269,
			65510,
			65231&L_{0_{2}a_{2}}\\\hline
		\end{array}
		\]
		\caption{Verstraete type of forms in the third secant variety}\label{TabLM}
	\end{table}
	

\begin{rem}\rm
 Our algorithm is based on a discussion on the roots of the quartics $\mathcal{Q}_1, \mathcal{Q}_2$ and $\mathcal{Q}_3$. It can also be seen using 
 the geometrical approach of Section \ref{geo}:
 \begin{enumerate}
\item If $\ket{\varphi}$ do not vanish the hyperdeterminant $\Delta$ then we are in cases 1.(a), 2.(a) and 3.(a).
\item If $\ket{\varphi}$ is a smooth point of $\Delta$, then we are in cases 1.(b), 2.(d) and 3.(b).
\item If $\ket{\varphi}$ is a smooth point of the cusp component ($\tau(X)^*)$ then we are in cases 1.(c) and 2.(e).
\item If $\ket{\varphi}$ is a smooth point of the cusp component of multiplicity $3$ ($X_{cusp,3}$), then we are in case 2.(c).
\item If $\ket{\varphi}$ is a smooth point of the node compotents, then we are in the case 2.(b).
\item If $\ket{\varphi}$ is a smooth point of the  node component of multiplicity $3$ ($X_{node,3}$), then we are in the case 3.(c).
\item Otherwise $\ket{\varphi}$ belongs to the nullcone.
 \end{enumerate}

\end{rem}
\begin{rem}\rm
 As shown in Ref\cite{CDGZ} two states of the same Verstraete form are SLOCC equivalent (up to a qubit permutation) if they take the same values on the four-qubit invariants. Thus
 the algorithm can be used to decide whether two given states are equivalent.
\end{rem}

\section{Conclusion}\label{conclu}
In our previous papers\cite{HLT,HLT2} we have proposed algorithmic methods based on invariants and covariants to identify the entanglement class of a given state  when the 
number of orbits 
is finite. This approach remains efficient for studying the nilpotent four-qubit states because the nullcone $\mathcal{N}\subset \PP(\mathcal{H})$ contains a finite number of SLOCC orbits.
However a covariant classification of $\PP(\mathcal{H})\diagdown \mathcal{N}$ is hopeless as we know that the parametrization of $\PP(\mathcal{H})$ by SLOCC orbits depends on parameters. 
Nevertheless the three quartics $\mathcal{Q}_1, \mathcal{Q}_2$ and $\mathcal{Q}_3$ obtained by the three natural embeddings of $\mathcal{H}$ in $\mathfrak{so}(8)$ lead to
a  stratification of the ambient space according to the configuration of their roots. This discussion has a natural geometric counterpart through the concept of dual variety: The existence of a zero root means the state belongs to the dual of one of the three 
embedding of $\PP^3\times \PP^3$ while the existence of a multiple root means that the state belongs to some singular locus of the hyperdeterminant. In the spirit of the earlier work of Miyake\cite{My} we 
pushed forward the investigation of the singular stratas of the hyperdeterminant. We showed the existence of six stratas whose general points correspond (up to a qubit permutation) to 
the six families of Verstraete {\em et al.} classification depending on parameters. The three families which do not depend on parameters correspond
to  stratas of the nullcone and were studied in Ref\cite{HLT2}. Thus, identifying the Verstraete form of a state is similar to finding to which strata 
this state belongs. 
This can be achieved by the use of invariants (of the three quartics) and covariants as explained in Section \ref{algorithme}. This algorithm, thanks to the  result of Chen {\em et al.}\cite{CDGZ}, 
can be used to determine whether 
 two given four-qubit states are SLOCC equivalent up to a permutation of  qubits.

\section*{Acknowledgement}
The authors would like to thank P\'eter L\'evay for sharing with them his ideas on the geometry of four-qubit invariants.

\appendix
\section{Demitesseract\label{demitesseract}}
\begin{figure}[!h]
\begin{center}
\includegraphics[width=4cm]{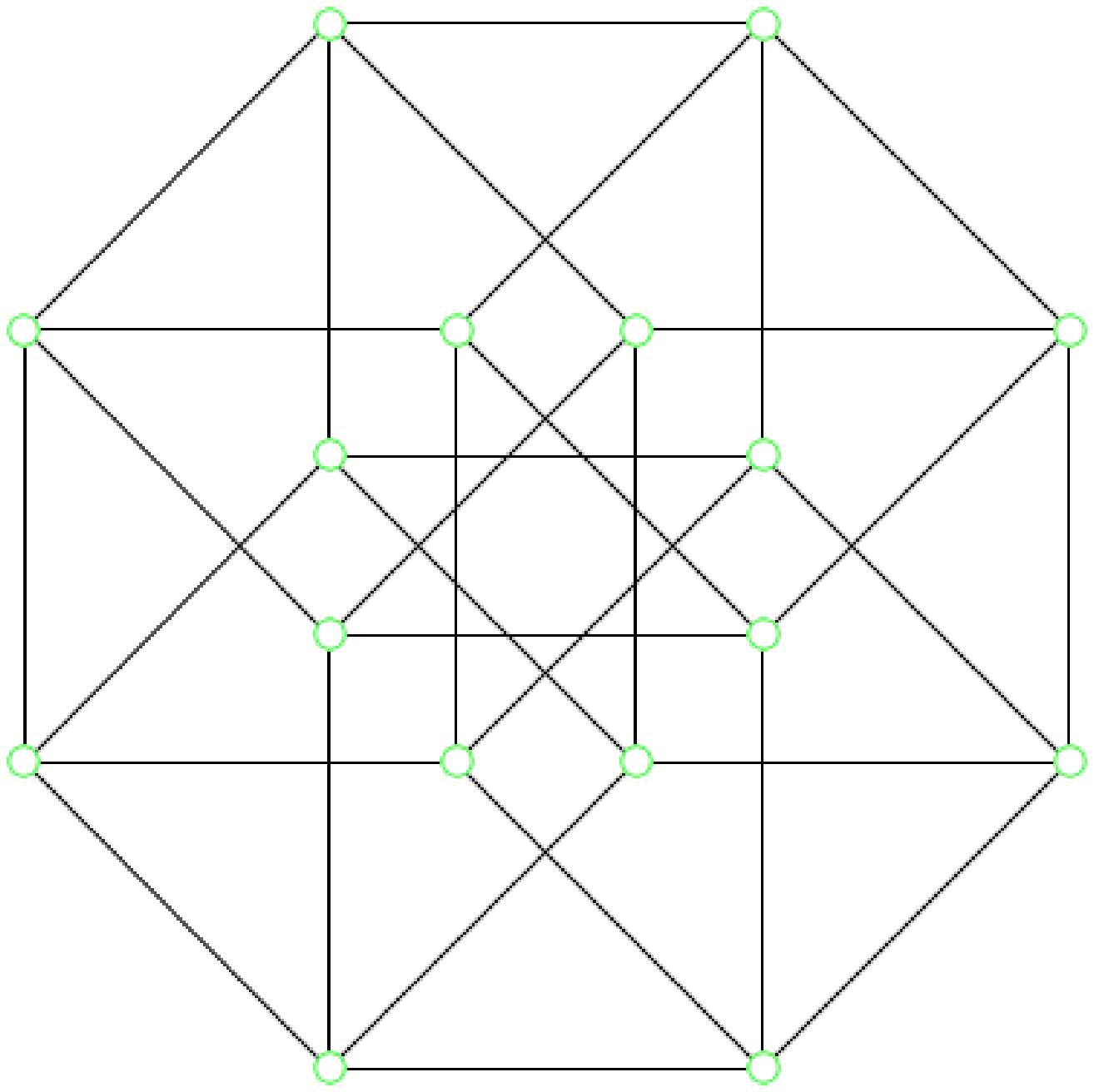} \includegraphics[width=4cm]{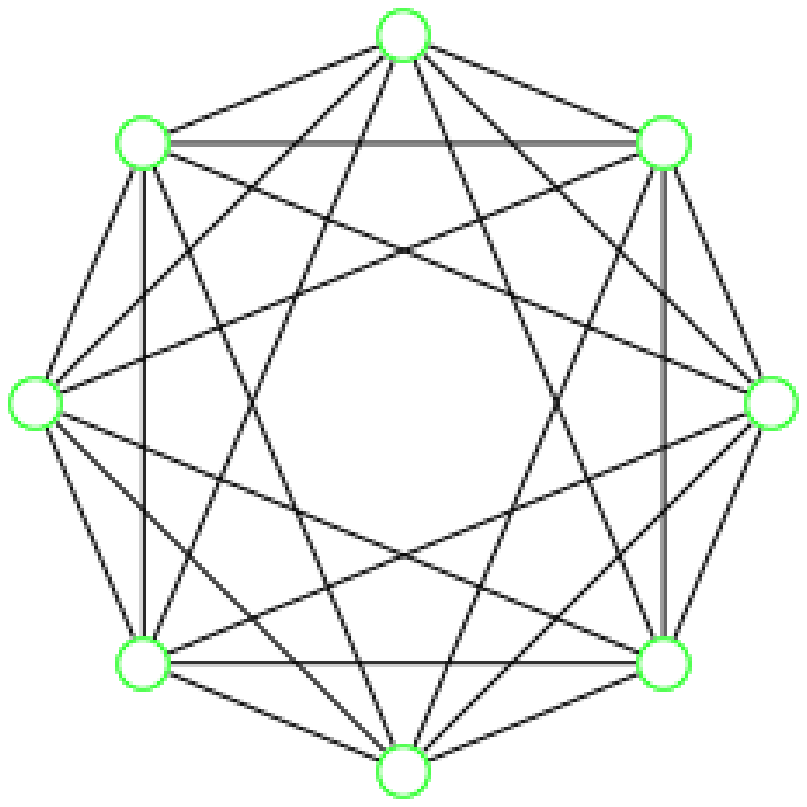} \includegraphics[width=4cm]{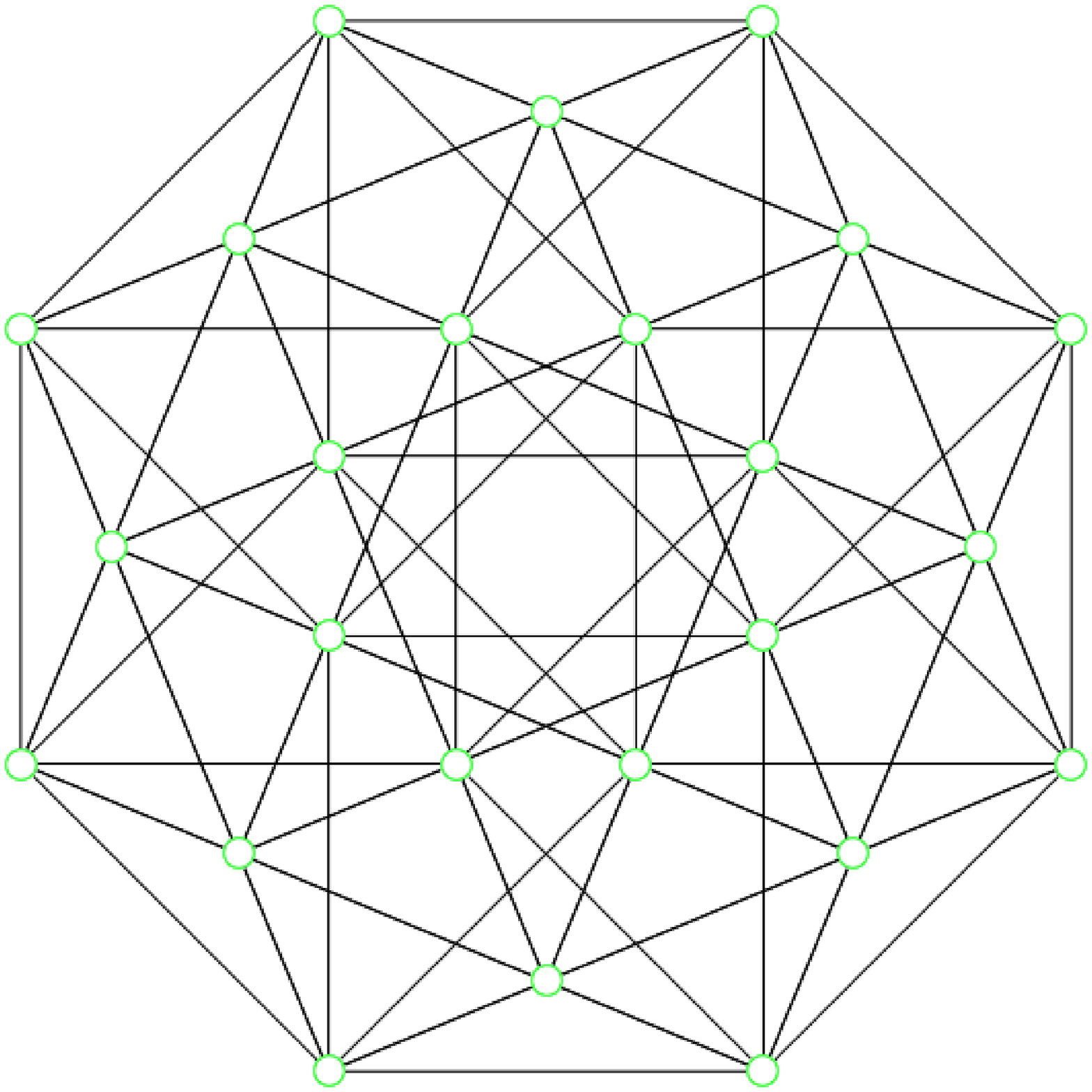}
\end{center}
\caption{Octogonal\cite{Cox24cell} projection of the tesseract, the 16-cell and the 24-cell\cite{CoxBook}. \label{4dpolytopes}}
\end{figure}

\begin{figure}[!h]
\begin{center}
\includegraphics[width=5cm]{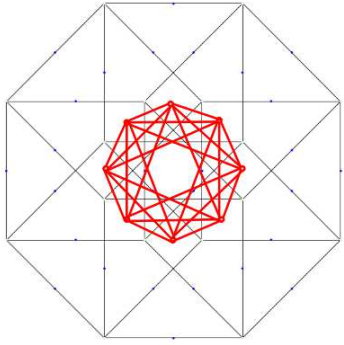}
\end{center}
\caption{A hexadecachoron (in red) whose vertices are the centers of the cells of a tesseract.\label{reciprocal}}
\end{figure}

The study of the higher-dimensional regular polytopes was pioneered by the Swiss mathematician Ludwig Sch\"afli in the middle of the 19th century, introducing Sch\"afli symbols which describe all tesselations of an $n$-sphere. The list of regular polytopes was extended to complex polytopes by Shephard in 1952. Readers interested in the subject may refer to an impressive series of books and papers written by Coxeter \cite{CoxBook,CoxComplex}.
Regular and semiregular polytopes form a family of geometrical objects whose symmetries are generated by mirrors. In dimension 4, the list of the (real) regular polytopes contains $6$ Figures: the $5$-cell (also called the $4$-simplex) which is the $4$-dimensional analogue of the tetrahedron, the tesseract (hypercube in dimension $4$), the hexadecachoron (also called $16$-cell) which is the $4$-dimensional analogue of the octahedron, the icositetrachoron (also called $24$-cell), the $120$-cell and the $600$-cell. In the paper, we use only three of them : the tesseract ($16$ vertices, $32$ edges, $24$ faces, and $8$ cells), the $16$-cell ($8$ vertices, $24$ edges, $32$ faces, and $16$ cells) and the $24$-cell ($24$ vertices, $96$ edges, $96$ faces, and $24$ cells), see Figure \ref{4dpolytopes}. To each polytope, one can associate a dual polytope whose vertices are constructed from the center of its cells. The tesseract and the $16$-cell are dual to each other (see Figure \ref{reciprocal}) and the $24$-cell is 
self dual. The middle of the edges of a $16$-cell or the center of the faces of a tesseract are the vertices of a $24$-cell. The finite reflection groups associated with their respective mirrors are $B_4$ for the tesseract and the hypercube, and $F_4$ for the $24$-cell. Notice that $B_4$ is a subgroup of $F_4$.  

\begin{figure}[!h]
\begin{center}
\includegraphics[width=5cm]{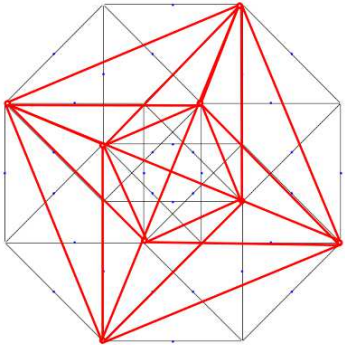}
\end{center}
\caption{Projection of a demitesseract  (in red) constructed on the vertices of a tesseract.\label{FigTesseract}}
\end{figure}

The demitesseract belongs to the family of demihypercubes, also called half measure polytopes and denoted by $h\gamma_n$, which are constructed from hypercubes by deleting half of the vertices and forming new facets in place of the deleted vertices \cite{CoxIII} (see Figure \ref{FigTesseract}). 
The demitesseract is the demihypercube in dimension $4$. As a polytope, the demitesseract is identical to the regular hexadecachoron.
Whilst the hexadecachoron and the demitesseract are identical as polytopes, their associated symmetries are different. Indeed, the reflection group of the hexadecachoron is $B_4$ that is the same of the tesseract. The reflections of the demitesseract  generates a subgroup of index $2$ of $B_4$ whose associated mirrors are \begin{equation}x_1=x_2,\ x_2=x_3,\ x_3=x_4,\mbox{ and }x_1+x_2=0.\end{equation} This group is generated by the transpositions $x_1\leftrightarrow x_2$, $x_2\leftrightarrow x_3$, and $x_3\leftrightarrow x_4$ together with an additional generator which transposes $x_1$ and $x_2$ while reversing the sign of both. So, clearly these reflections generate a group isomorphic to $D_4$. The normalized demitesseract has $8$ vertices, whose coordinates are $(1,1,1,1)$, $(1,1,-1,-1)$, $(1,-1,1,-1)$, $(-1,1,1,-1)$, $(1,-1,-1,1)$, $(-1,1,-1,1)$, $(-1,-1,1,1)$, and $(-1,-1,-1,-1)$, $24$ edges, $32$ triangular faces and $16$  tetrahedral cells. 

\section{Discussion on the quartics $Q_1$, $Q_2$ and $Q_3$\label{DiscRoots}}
First we focus the discussion on the number of zero roots of the quartic. We have to treat several cases.
\subsection{The three quartics have at least a zero root}
This implies that $L=M=0$ and this has been completely investigated in our previous paper\cite{HLT2}. The quartics are
\begin{equation}
Q_1=Q_2=Q_3=x^4-2Bx^3y+B^2x^2y^2+4D_{xy}xy^3.
\end{equation} 
If $D_{xy}=0$, the quartics have at least two zero roots and we obtain
\begin{equation}
Q_1=Q_2=Q_3=x^2(By-x)^2.
\end{equation}
So the only special case occurs when $B=0$ and corresponds to the nilpotent forms. The classification of nilpotent forms is well-known and is due to  Djokovic {\em et al.} \cite{CD,DLS} (see also the previous paper of the authors \cite{HLT2} for the geometrical interpretation). Note also that the case when $B\neq 0$ has
 been already algebraically and geometrically described \cite{HLT2}.\\
If $D_{xy}\neq 0$ then the discriminant is
\begin{equation}
\Delta=D_{xy}(27D_{xy}-B^3).
\end{equation}  
One has first to examine the case when  the roots of each quartic are distinct. This corresponds to $\Delta\neq 0$ and has been investigated in our previous paper\cite{HLT2} (no other cases appear in the present discussion).\\
If the quartics have nonzero multiple roots then the forms belong to the variety defined by
\begin{equation}
27D_{xy}-B^3=0.
\end{equation}
This variety have been also investigated in our previous paper\cite{HLT2}. Nevertheless the present discussion may refine the classification by regarding the other covariants of the quartics:
\begin{equation}
\begin{array}{rcl}
T&=&1152(D_{xy}x^6+D_{xy}Bx^5y+10D_{xy}^2x^3y^2-5D_{xy}^2Bx^2y^4+B^2D_{xy}^2xy^5+2D_{xy}^3y^6)\\
Hess&=&-12(B^2x^4-2(12D_{xy}+B^3)x^3y+B(12D_{xy}+B^3)x^2y^2+4B^2D_{xy}xy^3+12D_{xy}^2y^4)\\
I_2&=&2BD_{xy}+\frac1{12}H^4\\
I_3&=&-(D_{xy}^2+\frac1{216}B^6+\frac16B^3D_{xy}).
\end{array}
\end{equation}
Since $D_{xy}\neq 0$ the covariants $T$ and $Hess$ are clearly nonzero. According to Table \ref{RootQuart} the only case to investigate is $I_2=I_3=0$. But this implies $B=D_{xy}=0$ and this case has already been dealt with. It follows that all the interesting cases have been already investigated in our previous article \cite{HLT2}.
\subsection{Only one of the quartics has a zero roots}

\subsubsection{The quartic $Q_1$ has a zero root}
In this case, one has  $L=0$ and $M=-N$ (we will see that the other two cases are symmetrical). The quartics are
\begin{eqnarray}
Q_1=x^4-2Bx^3y+(B^2+4M)x^2y^2+4(D_{xy}-BM)xy^3,\\
Q_2=Q_3=x^4-2Bx^3y+(B^2-2M)x^2y^2-2(BM-2D_{xy})xy^3+M^2y^4.
\end{eqnarray}
The covariants reads
\begin{eqnarray}
I_2=2B(D_{xy}-BM)+\frac1{12}(B^2+4M)^2,\\
I_3=-(D_{xy}-BM)^2-\frac1{216}(B^2+4M)^3-\frac16B(B^2+4M)(D_{xy}-BM),
\end{eqnarray}
\begin{equation}
\begin{array}{rcl}
Hess(Q_1)&=&-12\, ( {{\it B}}^{2}-8\,{\it M} ) {x}^{4}+24\, ( {{
\it B}}^{3}-8\,{\it B}\,{\it M}+12\,{\it D_{xy}}) {x}^{3}y-12
\, ( 12\,{\it B}\,{\it D_{xy}}-4\,{{\it B}}^{2}{\it M}\\&&+{{\it B}
}^{4}+16\,{{\it M}}^{2} ) x^2{y}^{2}+48\, ( {{\it B}}
^{2}+4\,{\it M} )  ( -{\it D_{xy}}+{\it B}\,{\it M}
 ) x{y}^{3}-144\, ( -{\it D_{xy}}+{\it B}\,{\it M} ) 
^{2}{y}^{4},
\end{array}
\end{equation}
\begin{equation}
\begin{array}{rcl}
Hess(Q_2)&=&-12\, \left( {{\it B}}^{2}+4\,{\it M} \right) {x}^{4}+24\,( {{
\it B}}^{3}-8\,{\it B}\,{\it M}+12\,{\it D_{xy}} ) {x}^{3}y-12
\, ( -10\,{{\it B}}^{2}{\it M}+12\,{\it B}\,{\it D_{xy}}\\&&+{{\it B}}^{4}-8\,{{\it M}}^{2} ) x^2{y}^{2}+24\, ( {{\it B
}}^{3}{\it M}-2\,{{\it B}}^{2}{\it D_{xy}}-8\,{\it B}\,{{\it M}}^{2}
+4\,{\it M}\,{\it D_{xy}} ) x{y}^{3}\\&&-12\, ( {{\it B}}^{2}{{
\it M}}^{2}+4\,{{\it M}}^{3}-12\,{\it D_{xy}}\,{\it B}\,{\it M}+12\,
{{\it D_{xy}}}^{2}) {y}^{4},
\end{array}
\end{equation}
\begin{equation}
\begin{array}{rcl}
T(Q_1)&=&-1152(\,{\it D_{xy}}\,{x}^{6}- ( {\it B}\,{\it D_{xy}}+4\,{{
\it M}}^{2} ) {x}^{5}y+10\,{\it M}\, ( -{\it D_{xy}}+{
\it B}\,{\it M} ) x^4{y}^{2}\\&&-10\, ( -{\it D_{xy}}+{
\it B}\,{\it M} ) ^{2}{x}^{3}{y}^{3}+5\,{\it B}\, ( -
{\it D_{xy}}+{\it B}\,{\it M} ) ^{2}x^2{y}^{4}\\&&-\,
 ( {{\it B}}^{2}+4\,{\it M} )  ( -{\it D_{xy}}+{\it B
}\,{\it M} ) ^{2}x{y}^{5}+2\, ( -{\it D_{xy}}+{\it B}\,{
\it M} ) ^{3}{y}^{6}),
\end{array}\end{equation}
and
\begin{equation}
\begin{array}{rcl}
T(Q_2)&=&-1152\, ( {\it D_{xy}}-{\it B}\,{\it M} )  ( 
{x}^{6}
-{\it B}\,{x}^{5}y
+5\,{\it M}\,x^4{y}^{2}
-10\,{\it D_{xy}}\,x^3{y}^{3}\\&&
+ 5( \,{\it B}\,{\it D_{xy}}-\,{{\it M}}^{2} ) x^2{y}^{4}
- ( {{\it B}}^{2}{\it D_{xy}}-{\it B}\,{{\it M}}^{2}-2\,{\it M}\,{\it D_{xy}} ) x{y}^{5}
 \\&&+( {\it D_{xy}}\,{\it B}\,{\it M}-{{\it M}}^{3}+2\,{{\it D_{xy}}}^{2} ) {y}^{6} ) 
.\end{array}\end{equation}

According to Table \ref{RootQuart}, we have to investigate several cases.\\
{$\bullet$ \it The quartic $Q_1$ has a double zero root}\\
This case is identified by the equation $D_{xy}=BM$ and implies automatically $\Delta=0$. The quartics are
\begin{eqnarray}
Q_1=x^4-2Bx^3y+(B^2+4M)x^2y^2,\\
Q_2=Q_3=(-x^2+Bxy+y^2)^2.
\end{eqnarray}
The covariants simplify as
\begin{eqnarray}
I_2=\frac12(B^2+4M)^2,\\
I_3=\frac1{216}(B^2+4M)^3,\\
T(Q_1)=-1152(BMx^6-M(B^2+4M)x^5y),\\
Hess(Q_1)=-12(B^2-8M)x^4+24(B^2+4M)Bx^3y-12(B^2+4M)^2x^2y^2,\\
Hess(Q_2)=Hess(Q_3)=-12(B^2+4M)(Bxy+My^2-x^2)^2.
\end{eqnarray} 
Since $Q_2$ has two double roots, one has $T(Q_2)=0$. It remains to examine the following cases
\begin{enumerate}
\item If $Q_1$ has two simple distinct nonzero roots then $B^2+4M\neq 0$. It follows that $I_2, I_3\neq 0$ and $Q_2$ has no triple root. The only remaining case 
to consider is the case when $Q_2$ has a quadruple root. But this implies $B^2+4M=0$ and so it is not possible.
\item If $Q_1$ has a nonzero double root  then $T(Q_1)=0$ and this implies $M=D_{xy}=0$. This configuration cannot occur since it implies that $Q_2$ has a zero root.
\item If $Q_1$ has a triple zero root then $B^2+4M=0$ and $Hess(Q_2)=0$. Hence, $Q_2$ has a quadruple root.
\item If $Q_1$ has a quadruple zero root then $M=0$. So this configuration is not possible.
\end{enumerate} 
In conclusion, when $L=0$ and $M\neq 0$, we have only to investigate the variety given by $B^2+4M=0$.\\ \\
{$\bullet$ \it $Q_1$ has a nonzero double root and two simple roots}\\
Since the zero roots of $Q_1$ is not double, one has $D_{xy}\neq BM$ and then $T(Q_1), T(Q_2)\neq 0$. This prove that $Q_2$ has exactly one double root.\\
{$\bullet$ \it $Q_1$ has a triple nonzero root}\\
We have $D_{xy}\neq BM$ and $I_2=I_3=0$. The equation $I_2=I_3=0$ admits two solutions: $\{M=-\frac14B^2, D_{xy}=-\frac14B^3\}$ and $\{M=\frac1{12}B^2, D_{xy}=\frac1{108}B^3\}$. The first solution implies $D_{xy}=BM$, so it must be excluded.  From the second solution, we deduce
\begin{eqnarray}
T(Q_1)=-\frac{32}{2187}B^3(2By-3x)^6,\\
T(Q_2)=\frac4{2187}B^3(By-6x)^6,
Hess(Q_1)=-\frac4{81}B^2(2By-3x)^4,\\
Hess(Q_2)=-\frac1{81}B^2(By-6x)^4.
\end{eqnarray}  
The only special case is $B=0$ and that the form is nilpotent.\\
{\it $\bullet$ $Q_1$ has only simple roots}\\
In this case $Q_2$ has also only simple roots. This case corresponds to
\begin{equation}
\Delta=(B^3D_{xy}-B^2M^2-18D_{xy}BM+16M^3+27D_{xy})(BM-D_{xy})^2\neq 0.
\end{equation}
\subsubsection{$Q_2$ has a zero root}
In this case, one has  $M=0$ and $L=-N$. The quartics are
\begin{eqnarray}
Q_2=x^4-2Bx^3y+(B^2-4L)x^2y^2+4D_{xy}xy^3,\\
Q_1=Q_3=x^4-2Bx^3y+(B^2+2L)x^2y^2-2(BL-2D_{xy})xy^3+L^2y^4.
\end{eqnarray}
The covariants read
\begin{eqnarray}
I_2=3\left(\frac16B^2-\frac23L\right)^2+2BD_{xy},\\
I_3=-D_{xy}^2-\left(\frac16B^2-\frac23L\right)^3-B\left(\frac16B^2-\frac23L\right)D_{xy},
\end{eqnarray}
\begin{equation}
\begin{array}{rcl}
Hess(Q_2)&=&-12\, \left( {{\it B}}^{2}+8\,{\it L} \right) {x}^{4}+24\,( {{
\it B}}^{3}-4\,{\it B}\,{\it L}+12\,{\it D_{xy}} ) {x}^{3}y-12
\, ( -8\,{{\it B}}^{2}{\it L}+12\,{\it B}\,{\it D_{xy}}\\&&+{{\it B}}^{4}+16\,{{\it L}}^{2} ) x^2{y}^{2}
-48(B^2-4L)x{y}^{3}-144\,
{{\it D_{xy}}}^{2}) {y}^{4},
\end{array}
\end{equation}
\begin{equation}
\begin{array}{rcl}
Hess(Q_1)&=&-12\, \left( {{\it B}}^{2}-4\,{\it L} \right) {x}^{4}+24\,( {{
\it B}}^{3}-4\,{\it B}\,{\it L}+12\,{\it D_{xy}} ) {x}^{3}y-12
\, ( -2\,{{\it B}}^{2}{\it M}+12\,{\it B}\,{\it D_{xy}}\\&&+{{\it B}}^{4}-8\,{{\it L}}^{2} ) x^2{y}^{2}-24\, ( -{{\it B
}}^{3}{\it L}+2\,{{\it B}}^{2}{\it D_{xy}}+4\,{\it B}\,{{\it L}}^{2}
+4\,{\it L}\,{\it D_{xy}} ) x{y}^{3}\\&&-12\, ( {{\it B}}^{2}{{
\it L}}^{2}-4\,{{\it L}}^{3}-12\,{\it D_{xy}}\,{\it B}\,{\it L}+12\,
{{\it D_{xy}}}^{2}) {y}^{4},
\end{array}
\end{equation}
\begin{equation}
\begin{array}{rcl}
T(Q_2)&=&-1152(\,{\it D_{xy}}-BL)\,{x}^{6}- ( {\it B}\,{\it D_{xy}}+4\,{{
\it L}}^{2}-B^2L ) {x}^{5}y+10\,{\it L}{\it D_{xy}} x^4{y}^{2}\\&&-10\,D_{xy}^2{x}^{3}{y}^{3}+5\,{\it B}D_{xy}^2x^2{y}^{4}-\,
 ( {{\it B}}^{2}-4\,{\it L} )D_{xy}^2x{y}^{5}-2\, D_{xy}^{3}{y}^{6}),
\end{array}\end{equation}
and
\begin{equation}
\begin{array}{rcl}
T(Q_1)&=&-1152\,  {\it D_{xy}}  ( 
{x}^{6}
-{\it B}\,{x}^{5}y
-5\,{\it L}\,x^4{y}^{2}
-10\,({\it D_{xy}}-BL)\,x^3{y}^{3}\\&&
+ 5( \,{\it B}\,{\it D_{xy}}-B^2L-\,{{\it L}}^{2} ) x^2{y}^{4}
- ( {{\it B}}^{2}{\it D_{xy}}-B^3L+2LD_{xy}-3BL^2 ) x{y}^{5}
 \\&&-( B^2L^2-L^3+2D_{xy}^2-3BLD_{xy} ) {y}^{6} ) 
.\end{array}\end{equation}
Hence, the reasoning is very similar to the case $L=0$. Let us summarize it below.\\
According to table \ref{RootQuart}, we have to investigate several cases.\\
{$\bullet$ \it The quartic $Q_2$ has a double zero root}\\
The case is identified by the equation $D_{xy}=0$ and implies $\Delta=0$, $Q_1=Q_3=(x^2-Bxy+L^2y^2)^2$, $Q_2=x^2(x^2-2Bxy+(B^2-4L)y^2)$, $I_2=\frac1{12}(B^2-4L)^2$ and $I_3=-\frac1{216}(B^2-4L)^3$.
Hence, when $L\neq 0$ and $M=0$, we have only to investigate the variety given by $B^2-4L=0$. This corresponds the case when $Q_1$ has a quadruple root and $Q_2$ has a triple zero root.\\
{$\bullet$ \it $Q_2$ has a nonzero double root and two simple roots}\\
In this case, $D_{xy}\neq 0$ and so $T(Q_1), T(Q_2)\neq 0$. So $Q_1$ has exactly one double root.\\
{$\bullet$ \it $Q_2$ has a triple nonzero root}\\
The only solution of $I_2=I_3=M=0$ satisfying $D_{xy}\neq 0$ is $D_{xy}=-\frac2{27}B^3$ and $L=-\frac1{12}B^2$. The only special case is $B=0$ and implies the nilpotence of the form.\\
{$\bullet$ \it $Q_2$ has only simple roots}\\
In this case $Q_1$ has also four distinct roots. Furthermore, the hyperdeterminant factorizes as
\[
\Delta=D_{xy}^2(36BLD_{xy}+B^4L-B^3D_{xy}-27D_{xy}^2+16L^3-8B^2L^2).
\]

\subsubsection{$Q_3$ has a zero root}

In this case, one has  $N=0$ and $L=-M$. 
The quartics are
\begin{eqnarray}
Q_3=x^4-2Bx^3y+(B^2-4M)x^2y^2+4D_{xy}xy^3,\\
Q_1=Q_2=x^4-2Bx^3y+(B^2+2M)x^2y^2-2(BM-2D_{xy})xy^3+M^2y^4.
\end{eqnarray}
So it is deduced by substituting $L\leftrightarrow M$ and $Q_2\leftrightarrow Q_3$ in the previous discussion ($M=0$).
\subsection{The quartics have no zero root}
We have to investigate three cases.
\subsubsection{$Q_1$ has four simple roots}
This means that $\Delta\neq 0$ and then $Q_2$ and $Q_3$ have both four simple roots.
\subsubsection{$T(Q_1)=0$}
The equation $T(Q_1)=0$ a has two solutions: $M=D_{xy}=0$ and $N=D_{xy}=0$. In this two cases, one of the quartic has a zero root. In the same way, if $T(Q_2)=0$ or $T(Q_3)=0$ then one of the quartic has a zero root.
\subsubsection{$T(Q_1),T(Q_2),T(Q_3)\neq 0$}
Since $Hess=0$ implies $T=0$ (if a form has a quadruple root then it has two double roots which are equal), one has two cases to consider:
\begin{enumerate}
\item $I_2, I_3\neq 0$,
\item $I_2=I_3=0$.
\end{enumerate}

\section{Symmetries of the Verstraete forms}
\subsection{Permutations of the qubits}
Versraete {\em  et al.}\cite{VDMV} gave nine inequivalent normal forms for the four qubic forms. 
These forms were defined up to a permutation of qubits.
One of these forms $G_{abcd}$ has $4$ parameters and the set of all the $G_{abcd}$ defines a $4$-dimension subspace 
of the ambient space.
The five other forms, $L_{abc_{2}}, L_{a_{2}b_{2}}, L_{ab_{3}}, L_{a_{4}}$ and $L_{a_{2}0_{3\oplus\overline 1}}$
 have one to three parameters and define five affine subspaces.
The remaining three forms, $L_{0_{5\oplus\overline 3}}, L_{0_{7\oplus\overline 1}}$ and $L_{0_{3\oplus\overline 1}0_{3\oplus\overline 1}}$ are nilpotent.
For a given form $\varphi$, we will denote by $\varphi^{\sigma}$ the form obtained by applying the permutation $\sigma$ on the qubits. 
For the more generic form $G_{abcd}$, we obtain $5$ other non equivalent forms $G_{abcd}^{1423}$, $G_{abcd}^{1324}$, $G_{abcd}^{1243}$, $G_{abcd}^{1324}$ and $G_{abcd}^{1432}$.
Each of them is equivalent to a $G_{a'b'c'd'}$ for one of the following specializations (which are involutions):
\begin{itemize} 
\item $(a,b,-c,d)$, 
\item $(\frac{a+b-c+d}2,
\frac{a+b+c-d}2,\frac{-a+b+c+d}2,\frac{a-b+c+d}2)$,  
\item $(\frac{a+b-c+d}2,
\frac{a+b+c-d}2,\frac{a-b-c-d}2,\frac{-a+b-c-d}2)$, 
 \item$(\frac{a+b+c+d}2,
\frac{a+b-c-d}2,\frac{a-b-c+d}2,\frac{-a+b-c+d}2)$,
 \item and $(\frac{a+b+c+d}2,
\frac{a+b-c-d}2,\frac{a-b+c-d}2,\frac{a-b-c+d}2)$.
\end{itemize}
The permutations of $L_{abc_{2}}$ split into $6$ families with representatives $L_{abc_{2}}$,
$L^{1342}_{abc_{2}}$,
$L_{abc_{2}}^{2431}$,
$L_{abc_{2}}^{1432}$,
$L_{abc_{2}}^{2341}$,
and
$L_{abc_{2}}^{2143}$.
Furthermore, we have
\[L^{1324}_{\left(\frac{a+b}2+c,\frac{a+b}2-c,\frac{a-b}2\right)}\sim L^{1243}_{abcc},\
L^{2431}_{\left(\frac{a+b}2+c,\frac{a+b}2-c,\frac{a-b}2\right)}\sim L^{2134}_{abcc},\]
\[L^{1432}_{\left(\frac{a+b}2+c,\frac{a+b}2-c,\frac{a-b}2\right)}\sim L^{1423}_{abcc},\ 
 L^{1234}_{\left(\frac{a+b}2+c,\frac{a+b}2-c,\frac{a-b}2\right)}\sim L^{1324}_{abcc},\]
\[L^{2341}_{\left(\frac{a+b}2+c,\frac{a+b}2-c,\frac{a-b}2\right)}\sim L^{2314}_{abcc},\ 
L^{2143}_{\left(\frac{a+b}2+c,\frac{a+b}2-c,\frac{a-b}2\right)}\sim L^{2413}_{abcc}.\]
There are also $6$ different families of permutations of $L_{ab_{3}}$ whose representatives are $L^{\sigma}_{ab_{3}}$ 
for $\sigma\in\{1234,1423,3421,1342,3412\}$.
The permutations of $L_{a_{2}b_{2}}$ generate $12$ non equivalent families $L^{\sigma}_{a_{2}b_{2}}$ for $\sigma\in\{1234,2314,2413,1324,1243,2134,2431,2143,3241,3124,3214\}$.
The permutations $L_{a_{4}}^{\sigma}$ for $\sigma\in\S_{4}$ are pairwise nonequivalent. Finally, the permutations of $L_{a_{2}0_{3\oplus\overline 1}}$ give $4$ nonequivalent families 
$L_{a_{2}0_{3\oplus\overline 1}}^{1234}$, $L_{a_{2}0_{3\oplus\overline 1}}^{2134}$, $L_{a_{2}0_{3\oplus\overline 1}}^{2341}$, and $L_{a_{2}0_{3\oplus\overline 1}}^{2314}$.

If we send each parameter to zero, the permutations of the Verstraete forms specialize to nilpotent orbits. 
To each form corresponds one of the strata defined in our previous paper \cite{HLT2} see Table \ref{TVN}.
\begin{table}
	\begin{tabular}{|c|c|}
		\hline Forms& Strata\\\hline
		$G_{abcd}$& $Gr_{0}$\\
		$L_{abc_{2}}$& $Gr_{1}$\\
		$L_{a_{2}b_{2}}$& $Gr_{2}$\\
		$L_{a_{2}0_{3\oplus\overline 1}}$& $Gr_{3}$\\
		$L_{0_{3\oplus\overline1}0_{3\oplus\overline1}}$& $Gr_{4}$\\
		$L_{ab_{3}}$& $Gr_{5}$\\
		$L_{a_{4}}$& $Gr_{6}$\\
		$L_{0_{5\oplus\overline 3}}$& $Gr_{7}$\\
		$L_{0_{7\oplus\overline 1}}$& $Gr_{8}$\\\hline
	\end{tabular}
\caption{Correspondence between Verstraete forms and nilpotent strata}\label{TVN}\end{table}
\subsection{Quadrics again}
In a general setting, permuting the qubits in a form induces a permutations on the quadrics $Q_{1}, Q_{2}, Q_{3}$.
 Let us denote $\mathcal Q(\varphi)=\{Q_{1}(\varphi),Q_{2}(\varphi),Q_{3}(\varphi)\}$.
We notice also that any of the quadrics in the Verstraete forms can be written as $Q_{1}(G_{abcd})$
for some specialization of the parameters $a$, $b$ and $c$.  Indeed we have
\[\mathcal Q(G_{abcd})=
\left\{Q_{1}\left(G_{abcd}\right), Q_{1}\left(G_{\frac{a+c-d-b}2\frac{a+d-b-c}2\frac{a+b-c-d}2\frac{a+b+c+d}2}\right){}
	Q_{1}\left(G_{\frac{a+b+c-d}2\frac{a+b-c+d}2\frac{a-b+c+d}2\frac{-a+b+c+d}2}\right)\right\}.\]
	The remaining values are summarized in Table \ref{TQF}.
%

\begin{table}
	\[\begin{array}{|c|c|c|c|c|c|c|}
	\hline \mbox{Forms}&\mathcal Q (G_{abcd}) &\mathcal Q (L_{abc_{2}})  &\mathcal Q (L_{ab_{3}}) &\mathcal Q (L_{a_{2}b_{2}}) & \mathcal Q  (L_{a_{4}})&\mathcal Q (L_{a_{2}0_{3\oplus\overline 1}})  \\\hline
	G_{abcd}  & \mathcal Q (G_{abcd}) & & & & & \\
	L_{abc_{2}}& \mathcal Q\left(G_{abcc}\right)& & & & & \\
	L_{ab_{3}}& \mathcal Q\left(G_{aaab}\right)&\mathcal{Q}(L_{aba}) & & & & \\
	 L_{a_{2}b_{2}}&\mathcal Q\left(G_{aabb}\right)&\mathcal Q (L_{aab}) & & & & \\
	 L_{a_{4}}& \mathcal Q\left(G_{aaaa}\right)& \mathcal Q (L_{aaa}) & \mathcal Q (L_{aa_3}) & & & \\
	 L_{a_{2}0_{3\oplus\overline 1}}&\mathcal Q\left(G_{aa00}\right)& \mathcal{Q} (L_{aa0})& \mathcal Q (L_{a_20_2}) & & & \\
	 \hline
	 L_{0_{5\oplus\overline 3}}&\mathcal Q\left(G_{0000}\right)& \mathcal{Q}(L_{000})& \mathcal{Q}(L_{0_20_2}) & \mathcal{Q} (L_{00_3}) &\mathcal{Q}(L_{0_4}) & \mathcal{Q} (L_{0_20_{3\oplus \overline{1}}}) \\
	 L_{0_{7\oplus\overline 1}}&\mathcal Q\left(G_{0000}\right)&\mathcal{Q}(L_{000})& \mathcal{Q}(L_{0_20_2}) & \mathcal{Q} (L_{00_3}) &\mathcal{Q}(L_{0_4}) & \mathcal{Q} (L_{0_20_{3\oplus \overline{1}}}) \\
	 L_{0_{3\oplus\overline1}0_{3\oplus\overline1}}&\mathcal Q\left(G_{0000}\right)& \mathcal{Q}(L_{000})& \mathcal{Q}(L_{0_20_2}) & \mathcal{Q} (L_{00_3}) &\mathcal{Q}(L_{0_4}) & \mathcal{Q} (L_{0_20_{3\oplus \overline{1}}}) \\\hline
	\end{array}	
	\]
	
	\caption{Values of $\mathcal{Q}$ for Verstraete forms}\label{TQF}
\end{table}

\end{document}